\documentclass[conference]{IEEEtran}
\IEEEoverridecommandlockouts
\usepackage{cite}
\usepackage{amsmath,amssymb,amsfonts}
\usepackage{algorithmic}
\usepackage{graphicx}
\usepackage{textcomp}
\usepackage{xcolor}

\def\BibTeX{{\rm B\kern-.05em{\sc i\kern-.025em b}\kern-.08em
    T\kern-.1667em\lower.7ex\hbox{E}\kern-.125emX}}

\def\ve{\varepsilon}
\def\G{\Gamma}
\def\no{\nonumber}
\def\rar{\rightarrow}
\def\dis{\displaystyle}
\def\le{\left(}
\def\ri{\right)}
\def\res{\mathop{{\rm Res}}\limits}

\newtheorem{theorem}{\bf Theorem}[section]

\begin{document}

\title{Inverse Laplace and Mellin integral transforms modified for use in quantum  communications  \\

\thanks{The work of G.A. was supported in part by the joint DAAD/CONICYT scholarship 2015/57144001. 
The work of I.K. was supported in part by Fondecyt (Chile) Grants Nos. 1040368, 1050512 and 1121030, by DIUBB (Chile) Grant Nos.  125009,  GI 153209/C  and GI 152606/VC.}  
}

\author{
	\IEEEauthorblockN{Gustavo \'Alvarez \IEEEauthorrefmark{1} and  Igor Kondrashuk\IEEEauthorrefmark{4}}
\IEEEauthorblockA{\IEEEauthorrefmark{1} \textit{ II. Institut f\"ur Theoretische Physik, Universit\"at Hamburg},  \\
Luruper Chaussee 149,  22761 Hamburg, Germany \\
Email: gustavo.alvarez@desy.de}
\IEEEauthorblockA{\IEEEauthorrefmark{4}\textit{Grupo de Matem\'atica Aplicada \& Grupo de F\'isica de Altas Energ\'ias  \& Centro de Ciencias  
         Exactas} \\  
  \textit{   \& Departamento de Ciencias B\'asicas, Universidad del B\'io-B\'io}, \\
         Campus Fernando May, Av. Andres Bello 720, Casilla 447, 
         3780227  Chill\'an, Chile \\
	Email: igor.kondrashuk@gmail.com}
}

\maketitle

\begin{abstract}
Integral transformations are useful mathematical tool to work out signals and  wave-packets in electronic devices.  They may be used in software protocols.   
Necessary knowledge may come from quantum field theory, in particular from quantum chromodynamics, in which the optic theorem and the renormalization group equation 
can be solved by a unique contour integral written in two different "dual" ways related between themselves by a complex map in the complex plane of  Mellin variable.      
The inverse integral transformation should be modified to be applied for these contour integral solutions.  These modified inverse transformations may be  used in security protocols for quantum computers. Here we do a brief review of the basic integral transforms and propose their modification for the extended domains. 
\end{abstract}


\section{Introduction} \label{IT}

Integral transformations are applied to  process  signals and wave-packets.  Fourier transformation could be a good example.   Electrical devices were constructed to carry out analog signals  based on their Fourier transforms.  Functions invariant with respect to Fourier transformations may be  of special interest for quantum electronics due to this invariance 
\cite{Kondrashuk:2008ec,Kondrashuk:2008xq,Allendes:2009bd,Kondrashuk:2009us}.    Such Fourier-invariant functions have applications to statistics and machine learning \cite{Zast,Faouzi,convergence}. 
 In this paper other integral transformations are considered.  We pay attention to Laplace transforms and to Mellin moments.  The Mellin moments are equivalent to Laplace transforms and are defined in the real domain $[0,1]$ 
of the transformed functions. The inverse transformation  of a Mellin moment of a function recovers the function in this domain   $[0,1]$.  

Solutions to integro-differential equations (IDEs)  in quantum field theory may be represented  in terms of contour integrals in the complex plane of a variable of the Mellin moment  in which variables of functions, that solve these IDEs, enter in arguments of these contour integrals   \cite{Alvarez:2016juq,Kondrashuk:2019cwi,Alvarez:2019eaa,Kondrashuk:2025ajb}. Some of these variables run  in the real domain $[0,1]$  while other variables run in the real domain $[0,\infty[$.   By making a complex diffeomorphism in the complex plane of the Mellin variable,  such contour integrals  can be transformed to other contour integrals which efficiently solve IDEs of another type 
 \cite{Alvarez:2016juq,Kondrashuk:2019cwi,Alvarez:2019eaa,Kondrashuk:2025ajb}.  This requires
modification of the inverse transformations for Mellin moments in order to include the values of arguments  in the domain  $[0,\infty[$.   Roughly speaking, 
we take the Mellin moments in the domain  $[0,1]$     but the inverse transformation should be taken for the argument belonging to the domain   $[0,\infty[$.  
The construction obtained in the result of this approach in the present paper is not a modification of the inverse Mellin transformation  because there is  only one 
pole in the complex plane of the Mellin variable or the transform does not meet the criteria of asymptotic behaviour in the complex plane of the Mellin variable in the case of many poles in the plane \cite{Gonzalez:2021vqh}.
 The Mellin transformation  is taken over the real domain    $[0,\infty[$.  

 The modified  inverse transformations are necessary to represent  optic theorem as a Schrodinger equation, which is the main tool to  study quantum communication processes for quantum computers \cite{Kondrashuk:2025ajb}.  In  the next Sections we collect the necessary formulae for various integral transforms  and modify the inverse  Laplace transformations and  the inverse transformations of Mellin moments.

\section{Mellin transform} \label{MTB}

The Mellin transformation will not be modified in this article, however the basic tricks that we show in this Section will play an important role in order to extend inverse transformations
of Mellin moments. Such a generalization is necessary because we need to work with the extended domain $[0,\infty[.$  This extended domain is used in the  Dokshitzer-Gribov-Lipatov-Altarelli-Parisi  (DGLAP)  IDE  \cite{Alvarez:2016juq,Kondrashuk:2019cwi,Kondrashuk:2025ajb}  where we have to work with the 
Mellin moment with respect to the momentum transfer variable that runs in this extended domain  $[0,\infty[$ in the inverse transformation of Mellin moments.   
The essential part of this Section has already been published in   \cite{Alvarez:2016juq} and  \cite{Gonzalez:2021vqh}. We reproduce it for convenience of the reader
to compare with the content of Sections \ref{LTB} and \ref{MMB}.  

We start with a brief review of the Mellin transformation. The constructions that appear here, in the Mellin transformation, will  be used in Sections \ref{LTB} and \ref{MMB} dedicated 
to Laplace transforms and to  Mellin moments, respectively.  We define Mellin transform as 
\begin{eqnarray} \label{MT}
MT[f(x),x](z) = \int\limits_0^{\infty} x^{z-1}f(x)~dx, 
\end{eqnarray}
in which the arguments in the brackets on the l.h.s. stand for the transforming function $f(x)$  and the integration variable $x$ of  this integral transformation. The inverse Mellin transformation is 
\begin{eqnarray} \label{MT Cauchy} 
f(x) = \frac{1}{2\pi i}\int\limits_{c-i\infty}^{c + i\infty}x^{-z} MT[f(x),x](z) ~dz, ~~~ x \in [0,\infty[.
\end{eqnarray}
The position point $c$ of the  vertical line  of the integration contour 
in the complex plane must be in the vertical strip $c_1 < c < c_2,$ the borders of the strip are defined by the condition that two integrals 
\begin{eqnarray} \label{holom} 
\int\limits_0^{1} x^{c_1-1}f(x)~dx  &  {\rm and } &  \int\limits_1^{\infty} x^{c_2-1}f(x)~dx
\end{eqnarray}
must be finite.  This means that
\begin{eqnarray*}
|f(x)| < 1/x^{c_1} \quad {\rm when} \; x \to +0, \\
  |f(x)|< 1/x^{c_2} \quad {\rm when} \; x \to + \infty .
\end{eqnarray*}
The conditions (\ref{holom}) mean that the Mellin transform $MT[f(x),x](z)$ is holomorphic in the strip $c_1 <  {\rm Re}~z  < c_2.$   Indeed, if the Mellin transformation  (\ref{MT}) exist 
for any $z$ from the strip $c_1 <  {\rm Re}~z  < c_2,$  then  integrability conditions (\ref{holom}) are valid, and $MT[f(x),x](z)$ is holomorphic in the same strip $c_1 < {\rm Re}~z < c_2$ because all the poles are outside
the strip due to these integrability conditions (\ref{holom}).

Should the contour in  Eq.(\ref{MT Cauchy}) be closed to the left complex infinity or to the right complex infinity  
depends on the explicit asymptotic behaviour of the Mellin transform  $MT[f(x),x](z)$ at the complex infinity.
We close to the left if the left complex infinity does not contribute 
and we close to the right if the right complex infinity does not contribute \footnote{In comparison, in the Mellin-Barnes transformation 
we choose to which infinity the contour should be closed by taking into account 
the absolute value of $x$ in  (\ref{MT Cauchy}) because the MB transform has already an established structure 
in a form of fractions of the Euler. $\G$ functions \cite{Allendes:2009bd,Allendes:2012mr,Gonzalez:2012gu,Gonzalez:2012wk,Kniehl:2013dma,Gonzalez:2016pgx,Gonzalez:2018gch,Gonzalez:2019mdu,Diaz:2024tdp}.  
However, MB transformation is only a particular case of Mellin transformation.}.
Under this condition the original function $f(x)$ may be reproduced  via calculation of the residues by the Cauchy formula.

One of the simplest examples of the Mellin transformation is
\begin{eqnarray*} 
\G(z) = \int\limits_0^{\infty} e^{-x}x^{z-1}~dx,   \\
   e^{-x} = \frac{1}{2\pi i}\int\limits_{c-i\infty}^{c + i\infty}x^{-z} \G(z) ~dz.
\end{eqnarray*}
The contour in the complex plane is the straight vertical line with ${\rm Re}~z = c$ is in the strip $0 < c < A,$ where $A$ is a real 
positive number, the contour must be closed to the left infinity.

The couple of Eqs.(\ref{MT}-\ref{MT Cauchy}) may be proven. The proof  is already known for more than a century and may be found  in any textbook dedicated to the theory of complex variable, but we need
to reproduce it here because a similar construction will be used in the generalization of the contour for the inverse transformation in Sections \ref{LTB} and \ref{MMB} dedicated to the Laplace transform and 
to the Mellin moment.  First, we write up the direct transformation proof  using significantly that  $MT[f(x),x](z)$ is holomorphic in the strip $c_1 <  {\rm Re}~z  < c_2,$  while $\delta$ in all this paper  stands for infinitesimally small real positive number, 
\begin{eqnarray} \label{MT direct}
MT[f(x),x](z) = \int\limits_0^{\infty} x^{z-1}f(x)~dx \no\\
 =  \frac{1}{2\pi i}\int\limits_0^{\infty} x^{z-1} dx\int\limits_{c-i\infty}^{c + i\infty}x^{-\omega} MT[f(x),x](\omega) ~d\omega  \\
= \frac{1}{2\pi i}\int\limits_0^1 x^{z-1} dx\int\limits_{c-i\infty}^{c + i\infty}x^{-\omega} MT[f(x),x](\omega) ~d\omega   \no\\
+ \frac{1}{2\pi i}\int\limits_1^\infty x^{z-1} dx\int\limits_{c-i\infty}^{c + i\infty}x^{-\omega} MT[f(x),x](\omega) ~d\omega \no\\
= \frac{1}{2\pi i}\int\limits_0^1 x^{z-1} dx\int\limits_{c_1-\delta-i\infty}^{c_1-\delta + i\infty}x^{-\omega} MT[f(x),x](\omega) ~d\omega \no\\
+  \frac{1}{2\pi i}\int\limits_1^\infty x^{z-1} dx\int\limits_{c_2+\delta-i\infty}^{c_2+\delta + i\infty}x^{-\omega} MT[f(x),x](\omega) ~d\omega  \no\\ 
= \frac{1}{2\pi i}\int\limits_{c_1-\delta-i\infty}^{c_1-\delta + i\infty}\frac{MT[f(x),x](\omega)}{z-\omega} ~d\omega  \no\\
- \frac{1}{2\pi i}\int\limits_{c_2+\delta-i\infty}^{c_2+\delta + i\infty}\frac{MT[f(x),x](\omega)}{z-\omega} ~d\omega \no\\
= \frac{1}{2\pi i}\int\limits_{c_1-\delta+i\infty}^{c_1-\delta - i\infty}\frac{MT[f(x),x](\omega)}{\omega-z} ~d\omega  \no\\ 
+  \frac{1}{2\pi i}\int\limits_{c_2+\delta-i\infty}^{c_2+\delta + i\infty}\frac{MT[f(x),x](\omega)}{\omega-z} ~d\omega \no\\
= \frac{1}{2\pi i}\oint\limits_{CR}\frac{MT[f(x),x](\omega)}{\omega-z} ~d\omega, \no
\end{eqnarray}
here $CR$ is a rectangular contour constructed from the two straight vertical lines from the formula (\ref{MT direct}) supplemented with two horizontal lines at the imaginary complex infinities of the strip $c_1 <  {\rm Re}~z  < c_2.$
The contour $CR$ is closed in the counterclockwise orientation. 

The inverse transformation proof is even simpler and may be used in order to define Dirac $\delta$-function, 
\begin{eqnarray} \label{MT inverse} 
f(x) = \frac{1}{2\pi i}\int\limits_{c-i\infty}^{c + i\infty}x^{-z} MT[f(x),x](z) ~dz  \no \\
= \frac{1}{2\pi i}\int\limits_{c-i\infty}^{c + i\infty}x^{-z}~dz\int\limits_0^{\infty} y^{z-1}f(y)~dy \no\\
= \int\limits_0^{\infty}\delta(\ln{(y/x)}) y^{-1}f(y)~dy = f(x),
\end{eqnarray}
this is valid due to the following relation 
\begin{eqnarray} \label{delta}
\frac{1}{2\pi i}  \int\limits_{c - i\infty}^{c + i\infty}e^{(x-y)z}  dz = \frac{1}{2\pi }  \int\limits_{-\infty}^{\infty}e^{(x-y)(c +i\tau)}  d\tau \no\\
=  \frac{e^{(x-y)c}}{2\pi }\int\limits_{-\infty}^{\infty}e^{i(x-y)\tau}  d\tau 
= e^{(x-y)c}\delta(x-y) \no\\
= \delta(x-y).
\end{eqnarray}

We may write many parameters (for example,  other complex variables), $\alpha_1,\dots,\alpha_n$  on which  the function $f$ may depend, and make an integral transformation with respect to only one of them
\begin{eqnarray} \label{multi-va} 
MT[f(x,\alpha_1,\dots,\alpha_n),x](z)  \no\\
= \int\limits_0^{\infty} x^{z-1}f(x,\alpha_1,\dots,\alpha_n)~dx. 
\end{eqnarray}

The integral on the r.h.s. of Eq.(\ref{MT}) may be seen as a sum of two integrals 
\begin{eqnarray} \label{MT-LT} 
\int\limits_0^{\infty} x^{z-1}f(x)~dx = \int\limits_0^{1} x^{z-1}f(x)~dx \no\\ 
+ \int\limits_1^{\infty} x^{z-1}f(x)~dx = \int\limits_{-\infty}^{0} e^{tz}f(e^t)~dt + \int\limits_{0}^{\infty} e^{tz}f(e^t)~dt  \no\\ 
=  \int\limits_{-\infty}^{\infty} e^{tz}f(e^t)~dt  = \int\limits_{-\infty}^{\infty} e^{-tz}f(e^{-t})~dt. \no
\end{eqnarray}

\section{Laplace transform}  \label{LTB}

Laplace transformation  is not widely used in quantum chromodynamics (QCD), the main tool is   Mellin moments. However, the Laplace transform of a function may be mapped to the Mellin moment 
of another function, 
there is one-to-one 
correspondence between these two transforms. Moreover, the Laplace transformation is more frequently used integral transformation in the theory of differential equations  than the Mellin moments are, this is why we start this Section with a brief review 
of the Laplace transformation and 
then modify a contour of the inverse Laplace transformation in Subsections \ref{LTE} and \ref{LTG}. The purpose of such a modification is to repeat then this trick for the Mellin moment in Section \ref{MMB} 
in order to extend the real domain of arguments of the inverse transformation for  
Mellin moment from the standard domain $[0,1]$ to the extended domain $[0,\infty[$  for the real argument  with respect to which the Mellin moment is taken.

We define Laplace transform of function $f(x)$ as 
\begin{eqnarray}  \label{LT}
L[f(x),x](z) = \int\limits_{0}^{\infty} e^{-xz}f(x)~dx. 
\end{eqnarray}
This transformation is defined only for the functions that have restricted exponential growth $a,$ that is $\left|f(x)\right| < Ae^{a x},$ $A$ 
is a real positive, in the right complex half-plane ${\rm Re}~z > a.$ In this case the inverse transformation is  
\begin{eqnarray}  \label{LT Cauchy} 
f(x) = \frac{1}{2\pi i}\int\limits_{a+\delta-i\infty}^{a+\delta+i\infty}e^{xz} L[f(x),x](z) ~dz, 
\end{eqnarray}  
where ${\rm Re} (z) = a + \delta$ and $\delta \to +0$. This means that the straight vertical line of the integration in the complex plane passes slightly to the right of the point $a$ which is usually called 
as critical exponent.  The Laplace transform $L[f(x),x](z)$ obtained by Eq. (\ref{LT}) for the  complex half-plane ${\rm Re}~z > a$ should be analytically continued to the whole complex plane $\mathbb{C}$
and due to restriction on the exponential growth, the analytically continued Laplace transform does not have poles to the right from the critical exponent, it has poles only to the left from the critical exponent,
that is, in the domain to which it has been continued analytically.  To prove the couple of Eqs. (\ref{LT} -\ref{LT Cauchy}),  we perform subsequent transformations and obtain identity 
\begin{eqnarray}  \label{LT direct}
L[f(x),x](z)  \no \\
= \frac{1}{2\pi i}\int\limits_{0}^{\infty} e^{-xz} ~dx \int\limits_{a+\delta-i\infty}^{a+\delta+i\infty}e^{xu} L[f(x),x](u) ~du  \no\\
= \frac{1}{2\pi i}\int\limits_{a+\delta-i\infty}^{a+\delta+i\infty}\frac{L[f(x),x](u)}{z-u}du =  L[f(x),x](z),
\end{eqnarray} 
where ${\rm Re}~z > a +\delta > a.$  The contour is closed to the right complex infinity, there is only one residue there.  We can close the contour to the left infinity too  and the result should be the same 
because analytically continued $L[f(x),x](u)$  has all its poles in the half-plane to the left from the vertical line which crosses the real axis at the point $a+\delta.$  However, the explicit proof
of this statement is quite long and we do not write it here. The inverse Laplace transformation can be checked as 
\begin{eqnarray} \label{LT inverse}
f(x) = \frac{1}{2\pi i}  \int\limits_{a+\delta-i\infty}^{a+\delta+i\infty}e^{xz}  L[f(x),x](z) dz \no\\
 = \frac{1}{2\pi i}  \int\limits_{a+\delta-i\infty}^{a+\delta+i\infty}e^{xz} \int\limits_{0}^{\infty} e^{-uz} f(u) ~du dz  \no\\
= \int\limits_{0}^{\infty} \delta(x-u) f(u) ~du  = f(x).
\end{eqnarray}
The last equality is due to Eq.(\ref{delta})  of Section \ref{MTB}.  \\  \\

\subsection{Inverse Laplace transformation of  $L[e^{-\gamma x}, x](z)$ in the domain $x\in]-\infty,\infty[$ }  \label{LTE}

For the future map to Mellin moments, we need to study the Laplace transformation of $e^{-\gamma x}$ with respect to the variable $x.$ 
The Laplace transformation can be applied in this case  because we have a critical exponent  for the growth of this  function, 
\begin{eqnarray*} 
 e^{-x{\rm Re}\gamma}  < e^{(-{\rm Re}\gamma + \delta)x}, ~~~ \delta > 0, ~~~~ x \in[0,\infty[. 
\end{eqnarray*}
This function   $e^{-\gamma x}$  may be used to show that the information about the values of function  $e^{-\gamma x}$  in the extended domain  $x \in \mathbb{R} $ may be obtained 
from its Laplace transform $L[e^{-\gamma x}, x](z)$ which is defined by Eq. (\ref{LT}) in all the present article.

Indeed, we consider first when $x$ is in the standard domain $[0,\infty[$ of the variable with respect to which the Laplace transformation should be done, 
\begin{eqnarray} \label{Laplace}  
L[e^{-\gamma x}, x](z)=   \int\limits_0^{\infty}  e^{-\gamma x} e^{-zx}dx  = \frac{1}{\gamma + z}.
\end{eqnarray}
The domain of complex variable $z$ for taking this Laplace transformation is ${\rm Re}(\gamma + z) > 0.$ In the complex plane of the variable 
$z$ the Laplace transform   $L[e^{-\gamma x}, x](z)$ may be analytically continued from the complex domain ${\rm Re}(\gamma + z) > 0$ to all the complex plane $\mathbb{C}$ of  the variable $z.$  
The inverse operation (\ref{LT Cauchy}) in this case is 
\begin{eqnarray} \label{x>0} 
  e^{-\gamma x} =  \frac{1}{2\pi i} \int\limits_{-{\rm Re} \gamma +\delta-i\infty}^{-{\rm Re} \gamma + \delta+i\infty}\frac{e^{zx}}{\gamma + z} ~dz,
\end{eqnarray} 
$\delta$ is a small positive real. Because $x \in [0,\infty[$ we must 
close the contour to the left in order to avoid any contribution from the complex infinity and use the Cauchy formula due to which only one 
residue at $z=-\gamma$ contributes.

However, just to map afterwards Laplace transforms to Mellin moments in Section \ref{MMB}, we need to extend the inverse Laplace transformation (\ref{LT Cauchy}) in such a way that it results in  
$e^{-\gamma x}$ not only for $x \in [0,\infty[$ but for any real value of $x.$    This generalization of the inverse Laplace transformation consists in 
modification of the contour in Eq. (\ref{LT Cauchy}) in a such way that we would be able to close it to the right complex infinity as well as to the left complex infinity. 
Roughly speaking, at present it makes a sense to close it to the left complex infinity for $x>0$  because if we close it to the right complex infinity for $x<0,$
which is not in the standard domain of the positive $x,$ 
the result would be zero due to the fact that the Laplace transform by its construction (\ref{LT}) does not have poles to the right from the critical exponent in the complex plane of the variable $z.$
For $x<0$ we may modify the contour to the form 
\begin{eqnarray} \label{x<0} 
e^{-\gamma x} = - \frac{1}{2\pi i} \int\limits_{-{\rm Re} \gamma - \delta-i\infty}^{-{\rm Re} \gamma - \delta+i\infty}\frac{e^{zx}}{\gamma + z} ~dz.
\end{eqnarray} 
The integral over this straight line is equal to zero for  the standard domain $x \in [0,\infty[$ because for this domain of $x$ 
we should close it to the left to avoid any  contribution of the complex infinity,  however  we do not have any contribution 
of the residues inside this contour.  In the case when $x \in ]-\infty,0]$ we need to close the contour to the right infinity in order to avoid 
any contribution of the complex infinity and in this case the only residue that contributes is the point $z=-\gamma.$

\begin{figure}[ht!] 
\centering\includegraphics[scale=0.4]{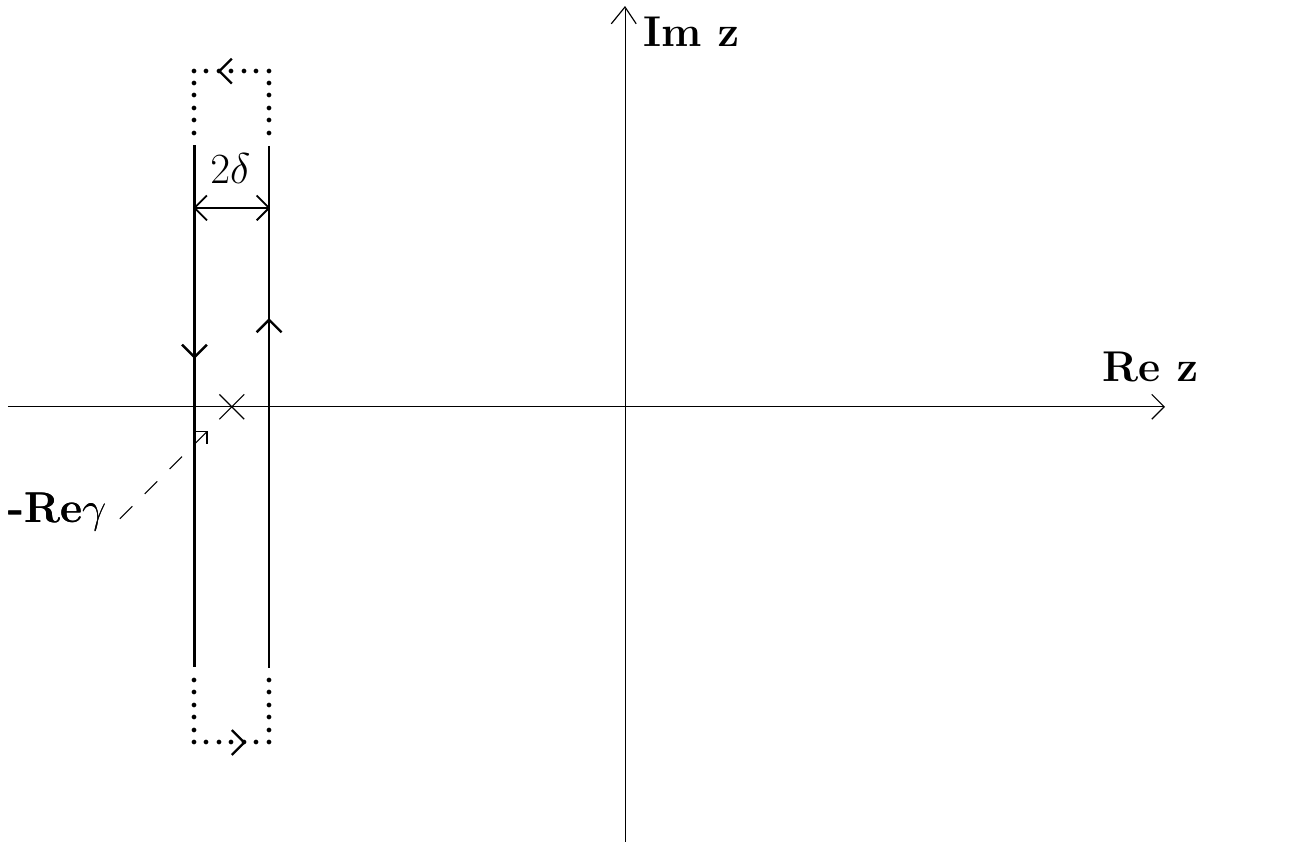}
\caption{\footnotesize  Contour $CR$  for the Laplace transform $L[e^{-\gamma x}, x](z)$}
\label{FLTE}
\end{figure}

To conclude, integral (\ref{x>0}) gives  $ e^{-\gamma x} $  for the domain  $x \in [0,\infty[$  and gives zero for the domain $x \in ]-\infty,0],$
while integral (\ref{x<0}) gives   $ e^{-\gamma x} $  for the domain $x \in ]-\infty,0]$ and gives zero for the  domain $x \in [0,\infty[.$ 
In the second case the negative sign appears due to clockwise orientation of the contour. Thus, we may write 
\begin{eqnarray}  \label{x} 
e^{-\gamma x}  = \frac{1}{2\pi i} \int\limits_{-{\rm Re} \gamma +\delta-i\infty}^{-{\rm Re} \gamma + \delta+i\infty}\frac{e^{zx}}{\gamma + z} ~dz  \no\\
- \frac{1}{2\pi i} \int\limits_{-{\rm Re} \gamma - \delta-i\infty}^{-{\rm Re} \gamma - \delta+i\infty}\frac{e^{zx}}{\gamma + z} ~dz 
\end{eqnarray}
and taking into account that $\delta \rar 0$ is infinitesimally small to avoid any contribution of the smallest sides of the rectangular 
shown in Fig. \ref{FLTE}, we may re-write this equality  as 
\begin{eqnarray} \label{LCR} 
 e^{-\gamma x}  = \frac{1}{2\pi i}\oint\limits_{CR} \frac{e^{zx}}{\gamma + z} ~dz,  ~~~~ x \in ]-\infty,\infty[, 
\end{eqnarray}
where we have a rectangular contour $CR$ which contains two straight vertical lines, one line crosses the real axis at the point at $z= -{\rm Re} \gamma +\delta$ and 
another crosses the real axis at the point $z = -{\rm Re} \gamma - \delta.$ In the right line the integration is performed from the negative imaginary 
infinity to the positive imaginary infinity, while for the left line we integrate down from the positive imaginary infinity to 
the negative imaginary infinity. This integration corresponds exactly to the counterclockwise orientation of the contour and by Cauchy formula 
corresponds to the only residue at the point $z= -\gamma$ in the complex plane of the variable $z.$  Strictly speaking,  it is not necessary to make the size of 
the smallest sides of the rectangular to be infinitesimally small in order to construct the extended contour (\ref{LCR}) of the inverse Laplace transformation,  
we may abandon this requirement and in Subsection \ref{LTG} we consider the rectangular contour with the finite size of the smallest sides. 
Eq.(\ref{LCR}) is valid for any positive real $\delta$ and  any real $x,$  that is, $x \in ]-\infty,\infty[.$ 
The extended inverse Laplace transformation  (\ref{LCR}) recovers the exponential  $e^{-\gamma x}$ in the extended domain $\mathbb{R}$ of all the real $x.$

We may repeat the direct  transformation   proof (\ref{LT direct}) we have used for the standard domain  $x \in [0,\infty[$ from the definition  (\ref{LT}) and apply it for the extended domain $x \in ]-\infty,\infty[,$ 
\begin{eqnarray}\label{LCR direct} 
\frac{1}{z+\gamma} = \int\limits_{0}^{\infty} e^{-\gamma x }e^{-zx} ~dx \no\\ 
 = \frac{1}{2\pi i}\int\limits_{0}^{\infty}e^{-zx}~dx\oint\limits_{CR} \frac{e^{\omega x}}{\gamma + \omega} ~d\omega \no\\
= \frac{1}{2\pi i}\oint\limits_{CR} \frac{1}{(\gamma + \omega)(z-\omega)} ~d\omega = \frac{1}{z+\gamma}.
\end{eqnarray}
Here the calculation of the residues may be done inside or outside the contour, the result will be the same. This may be proven that one of these two ways of calculation is equivalent to another. It is supposed
in Eq.(\ref{LCR direct}) that we are in the domain ${\rm Re}(\gamma + z) > 0$ of the complex plane $z.$
 
Also, we may repeat the inverse transformation proof   (\ref{LT inverse})  we have found for the standard domain $x \in [0,\infty[$ and apply it for the extended  domain $x \in~]-\infty,\infty[,$ 
\begin{eqnarray*}
e^{-\gamma x} =  \frac{1}{2\pi i}\oint\limits_{CR} \frac{e^{zx}}{\gamma + z} ~dz  \no\\
=  \frac{1}{2\pi i} \left[\int\limits_{-{\rm Re} \gamma +\delta-i\infty}^{-{\rm Re} \gamma + \delta+i\infty}e^{zx}~dz\int\limits_0^\infty  e^{-(\gamma + z)y} ~dy \right. \no\\
- \left. \int\limits_{-{\rm Re} \gamma - \delta+i\infty}^{-{\rm Re} \gamma - \delta-i\infty}e^{zx}~dz\int\limits_{-\infty}^0  e^{-(\gamma + z)y} ~dy \right]   \no
\end{eqnarray*}
\begin{eqnarray} \label{LCR inverse} 
= \int\limits_0^\infty \delta\le x -y \ri e^{-\gamma y} ~dy + \int\limits_{-\infty}^0 \delta\le x -y \ri  e^{-\gamma y} ~dy  \no\\
 =  \int\limits_{-\infty}^{\infty} \delta\le x-y\ri  e^{-\gamma y} ~dy. 
\end{eqnarray}
This chain of equalities is valid for any real  $x.$

Thus, in this Subsection we have generalized the standard inverse  Laplace transformation (\ref{x>0}) with the standard domain $x \in [0,\infty[$
to the extended inverse Laplace transformation  (\ref{LCR}) of the Laplace transform   $L[e^{-\gamma x}, x](z)$   which reproduces the exponential $e^{-\gamma x}$   for any $x$ from the extended 
domain  $x \in ]-\infty,\infty[.$

\subsection{Inverse Laplace transformation of   $L[f(x),x](z)$ in the domain $x \in  ]-\infty,\infty[$ } \label{LTG}

In Subsection \ref{LTE} we have considered an exponential function $f(x) = e^{-\gamma x},$ then we have calculated its Laplace transform  and have modified the inverse Laplace transformation in such a way that
it became possible to recover the original function  $e^{-\gamma x}$ in all  the range of real values of the variable $x$ by this extended inverse Laplace transformation (\ref{LCR}). The standard  inverse Laplace transformation 
(\ref{x>0})  recovered it for the standard domain $x>0$ only.   
In Section \ref{MMB}  we map the Laplace transform of the exponential function 
to the Mellin moment of the power-like function.  In this Subsection \ref{LTG} we write up an analog of Eq.(\ref{LCR}) in order to recover an arbitrary function $f(x)$ in the extended domain $x \in ]-\infty,\infty[,$ too,  
after making an extended inverse Laplace transformation of its Laplace transform $L[f(x),x](z)$  defined in Eq. (\ref{LT Cauchy}).
In the rest of this Subsection  we prove a possibility to modify the contour of the inverse Laplace transformation (\ref{LT Cauchy}) in order to reach this purpose.

\begin{figure}[hbt!!!] 
\centering\includegraphics[scale=0.4]{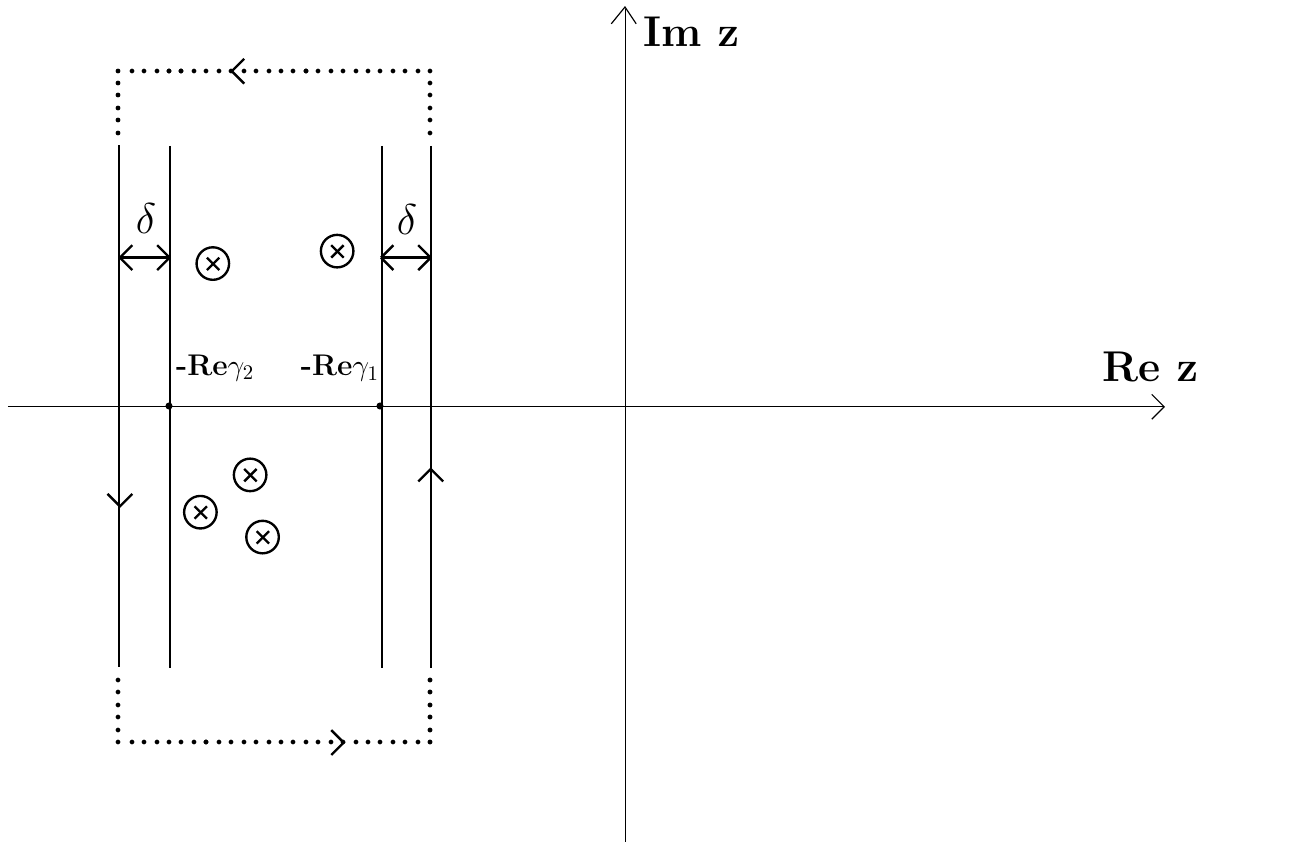}
\caption{\footnotesize  Contour $CR$ for the Laplace transform  $L[f(x),x](z)$ with poles inside }
\label{FLTG}
\end{figure}

First, we start with the standard inverse Laplace transformation (\ref{LT Cauchy}) for the standard domain $x \in [0,\infty[,$ and let $ -{\rm Re} \gamma_1$ be the critical exponent of the function 
$f(x),$  $L[f(x),x](z)$ is defined for ${\rm Re}~z  > - {\rm Re} \gamma_1.$ This means that all the poles are situated on the left hand side of the critical exponent, we have commented on this at the beginning of 
Section \ref{LTB}.  We continue analytically  $L[f(x),x](z)$ to the whole complex plane $z.$ There is a countable number of poles to the left from the vertical
line of the transformation (\ref{LT Cauchy}). This means that we may draw the second straight vertical line which passes a bit to the left of the leftmost pole in the complex plane $z,$  
and we construct  a rectangular contour  drawn in Fig. \ref{FLTG}.
All the poles of the Laplace  transform $L[f(x),x](z)$ in Eq. (\ref{LT Cauchy}) are inside  this contour in the complex plane $z.$  The second vertical line crosses the real axis of the plane $z$
at the point $-{\rm Re}~\gamma_2-\delta$ in Fig. \ref{FLTG}. This contour is closed to the rectangular form by two horizontal lines at the imaginary complex infinities in the strip  
$ - {\rm Re}~\gamma_2 < {\rm Re}~z  < -{\rm Re} \gamma_1.$

If $x>0$ we need to close each contour associated to every of these two straight vertical lines to the left complex infinity for both the vertical lines in order to avoid any 
contribution of the complex infinity on this contour, 
in such a case the right vertical line contributes with all the residues 
on the left hand side of it, and the left vertical line does not contribute at all because there is no residue on the left hand side of it by construction of this contour.  
This analysis repeats exactly the analysis done for 
Eq. (\ref{x>0}) which we have written in Subsection \ref{LTE} dedicated to the particular case of the exponential function.
If $x<0$ we need to close each contour associated to every of these two vertical lines to the right complex infinity for both the vertical lines to avoid any contribution of the complex infinity on this contour, 
in such a case the left vertical line contributes with all the residues 
on the right hand side of it, and the right vertical line does not contribute at all because there is no residue on the right hand side of it by construction of this contour.  This consideration is given in complete analogy
to Eq. (\ref{x<0}) of Subsection \ref{LTE}.  Whether both the contours associated to these vertical lines are closed to the right complex infinity or 
they are closed to the left complex infinity, the contribution of the residues will be the same and the result of this residue calculus will be $f(x).$  All this is written in complete analogy 
to the case of the exponential function considered in Subsection \ref{LTE} in Eq. (\ref{x}).

Thus, we may write 
\begin{eqnarray*}   
f(x)  =  \frac{1}{2\pi i} \int\limits_{-{\rm Re} \gamma_1 + \delta-i\infty}^{-{\rm Re} \gamma_1 + \delta+i\infty} e^{zx} L[f(x),x](z) ~dz  \\
- \frac{1}{2\pi i} \int\limits_{-{\rm Re} \gamma_2 -\delta-i\infty}^{-{\rm Re} \gamma_2 - \delta+i\infty} e^{zx} L[f(x),x](z)  ~dz 
\end{eqnarray*}
and then we may close the contour to the rectangular form at the complex imaginary infinities of the strip $-{\rm Re} \gamma_2 - \delta <  {\rm Re}~z  < -{\rm Re} \gamma_1 + \delta$  
as it is depicted in Fig.\ref{FLTG}.  As the result, we obtain complete analog of formula (\ref{LCR}) in which instead of the Laplace transform $L[e^{-\gamma x}, x](z) = 1/(\gamma + z)$  
the Laplace transform $L[f(x),x](z)$ of an arbitrary function $f(x)$ is written, 
\begin{eqnarray} \label{LCRG} 
f(x) = \frac{1}{2\pi i}\oint\limits_{CR} e^{zx} L[f(x),x](z)~dz,  ~~~~ x \in   ]-\infty,\infty[. 
\end{eqnarray}
Here it is worthy to mention again that the Laplace transform $L[f(x),x](z)$ is always defined by (\ref{LT}).  The rectangular contour $CR$  is depicted in Fig. \ref{FLTG}.
This contour contains two vertical lines, one line crosses the real axis at the point at $z= -{\rm Re} \gamma_1 +\delta$ and 
another crosses the real axis at the point $z = -{\rm Re} \gamma_2 - \delta.$ In the right line the integration is performed from the negative imaginary 
infinity to the positive imaginary infinity, while for the left line we integrate down from the positive imaginary infinity to 
the negative imaginary infinity. This integration corresponds exactly to the counterclockwise orientation of the contour and by Cauchy formula 
corresponds to the contribution of all the residues of   $L[f(x),x](z)$ in the strip  $-{\rm Re} \gamma_2 - \delta <  {\rm Re}~z  < -{\rm Re} \gamma_1 + \delta$   
in the complex plane of the variable $z.$   Eq.(\ref{LCRG}) is valid for any positive real $\delta$ and for all real $x$  that is, $x \in ]-\infty,\infty[.$ 
The extended inverse Laplace transformation  (\ref{LCRG}) recovers the function $f(x)$ in the extended domain  $] -\infty,\infty[.$

For the case of an arbitrary function $f(x)$ we may repeat the direct transformation proof (\ref{LCR direct}) of Subsection \ref{LTE} which we have found for the exponential function in the extended domain $x\in~]-\infty,\infty[,$
\begin{eqnarray} \label{LCRG direct}
L[f(x),x](z)  =  \int\limits_{0}^{\infty} f(x)e^{-zx} ~dx \no\\ 
= \frac{1}{2\pi i}\int\limits_{0}^{\infty}e^{-zx}~dx\oint\limits_{CR} e^{\omega x} L[f(x),x](\omega)  ~d\omega \\
= \frac{1}{2\pi i}\oint\limits_{CR} \frac{L[f(x),x](\omega)}{z-\omega} ~d\omega = L[f(x),x](z). \no
\end{eqnarray}
Here the calculation of the residues may be done inside or outside the contour. It may be proven that residue calculus inside the contour and the residue calculus outside the contour give the same results. This  formulae
suppose that we are in the standard domain ${\rm Re}(\gamma_1 + z) > 0$ of the complex plane of the Laplace transform variable $z.$

Also, we may repeat for an arbitrary $f(x)$ the inverse transformation proof  (\ref{LCR inverse})  which we have found in Subsection \ref{LTE} for the extended  domain $x\in~]-\infty,\infty[$ of the inverse Laplace 
transformation of $L[e^{-\gamma x}, x](z)$ that was based on the formula (\ref{LCR})
\begin{eqnarray} \label{LCRG inverse}
f(x)  =  \frac{1}{2\pi i}\oint\limits_{CR} e^{zx} L[f(x),x](z)~dz  \no\\
= \frac{1}{2\pi i} \left[\int\limits_{-{\rm Re} \gamma_1 +\delta-i\infty}^{-{\rm Re} \gamma_1 + \delta+i\infty}e^{zx}~L[f(x),x](z)~dz  \right. \no\\ 
+ \left. \int\limits_{-{\rm Re} \gamma_2 - \delta+i\infty}^{-{\rm Re} \gamma_2 - \delta-i\infty}e^{zx}~L[f(x),x](z)~dz \right]   \no
\end{eqnarray}
\begin{eqnarray}
= \frac{1}{2\pi i} \left[\int\limits_{-{\rm Re} \gamma_1 +\delta-i\infty}^{-{\rm Re} \gamma_1 + \delta+i\infty}e^{zx}~dz\int\limits_0^{\infty} f(u) e^{-zu} ~du   
\right. \no\\ 
- \left. \int\limits_{-{\rm Re} \gamma_2 - \delta+i\infty}^{-{\rm Re} \gamma_2 - \delta - i\infty}e^{zx}~dz\int\limits_{-\infty}^0 f(u) e^{-zu} ~du \right]   \no\\
= \int\limits_0^{\infty} \delta\le u-x\ri f(u)~du + \int\limits_{-\infty}^0 \delta\le u-x\ri  f(u) ~du  \no\\  
=  \int\limits_{-\infty}^{\infty} \delta\le u-x\ri f(u) ~du. 
\end{eqnarray}
This chain of equalities is valid for any real $x.$  The replacement of  $L[f(x),x](z)$ defined in (\ref{LT}) with a bit different expression $-\int_{-\infty}^0 f(u) e^{-zu} ~du$ in the second
integral of (\ref{LCRG inverse}) is justified by residue calculus.  This would be just a generalization of the case   $L[e^{-\gamma x}, x](z)$ considered in Eq. (\ref{LCR inverse}) 
to the Laplace transform of an arbitrary function $L[f(x), x](z)$ where   $-\gamma_1$ is the rightmost pole in the complex plane $z$ and $-\gamma_2$ is the leftmost pole in the complex plane $z$ of  $L[f(x),x](z).$
This means that the upper critical exponent is $-{\rm Re} \gamma_1 + \delta$
and the lower critical exponent is $-{\rm Re} \gamma_2 - \delta.$ For the case when $x \in~[0,\infty[$ that is, the variable $x$ is in the standard domain, we may reproduce inverse transformation proof
(\ref{LT inverse}) from this proof (\ref{LCRG inverse}) of the extended inverse Laplace transformation.

\subsection{Summarizing Laplace transforms}

Finally, at the end of this Section we would like to do three summarizing comments.  

\begin{itemize}
\item 
The domain of variable $x$ of  $f(x)$ should include the interval $x \in [0, \infty[,$ otherwise the Laplace transformation   (\ref{LT})
would be impossible to define. In brief, the  Laplace transform
is defined in the domain  ${\rm Re}~z > a,$ where $a$ is an index of the exponential growth of the function $f(x).$ 
In the standard inverse Laplace transformation (\ref{LT Cauchy}) the contour passed vertically in the complex plane $z$ at ${\rm Re}~z = a + \delta.$ 
Under this condition the  Laplace transform $L[f(x),x](z)$ does not have poles in the complex half-plane to the right from this vertical line  in the complex plane of variable $z.$  
\item 
An exponential upper bound for the dependence on the variable $x$ is the necessary condition for taking the Laplace transform of $f(x).$ 
In case if the lower bound for the exponential behaviour exist, the contour in the complex plane $z$ of the function  $L[f(x),x](z)$ would contain two vertical lines 
in such a manner that the left one is a bit to the left from the lower bound value, the right one is a bit to the right from the 
upper bound value on the real axis of the complex   plane of the variable $z.$   The contour of this type is shown in  Fig. \ref{FLTG}.
An example of such a type of the functions would be $e^{-\gamma_1 x}  + e^{-\gamma_2 x}.$ The number of residues inside the contour is countable.
The positions of the vertical lines of the contour depend on the bounds of the function $f(x)$ with respect 
to the variable $x.$ 
The function $L[f(x),x](z)$  may be continued analytically from the domain in which it is defined to all the complex plane of the variable $z.$  We may use this analytic continuation 
in order to recover the information for $f(x)$ for an arbitrary real domain of the 
variable $x$  from its Laplace transform  $L[f(x),x](z)$  (\ref{LT}).
\item 
To determine how the poles in the complex plane of variable $z$ are distributed, we need more information about the function $f(x).$  We have obtained 
the extended inverse Laplace transformation in Subsection \ref{LTG} in which the contour of this inverse transformation has a rectangular form. We may change the shape of   
the borders of this rectangular contour in any way under the condition that all the poles remain inside it. Then by Cauchy formula the result will be the same.
\end{itemize}

\subsection{Two theorems proven in Section \ref{LTB}}

\begin{theorem}\label{Laplace-1}
Let $\gamma \in \mathbb{C}$  and the formula 
\begin{eqnarray*} 
L[e^{-\gamma x}, x](z)=   \int\limits_0^{\infty}  e^{-\gamma x} e^{-zx}dx  = \frac{1}{\gamma + z}
\end{eqnarray*}
be the Laplace transform  of the function $e^{-\gamma x}$ defined in the complex plane of the variable $z$  for  ${\rm Re}(\gamma + z) > 0.$  Then, the
inverse integral transformation  
\begin{eqnarray*} 
 e^{-\gamma x}  = \frac{1}{2\pi i}\oint\limits_{C} \frac{e^{zx}}{\gamma + z} ~dz 
\end{eqnarray*}
recovers the function $e^{-\gamma x}$ for any  $x \in ]-\infty,\infty[.$ The contour $C$ is an arbitrary closed contour in the complex  $z$ plane such that  the point $z = -\gamma$ is inside the contour.
\end{theorem}

\begin{theorem}\label{Laplace-2}
Let functions $f_i(x)$ be an infinite or finite  set of smooth functions  of the real argument,
$f_i(x)\in C^{\infty}\le \mathbb{R}\ri.$ Let a function $f(x)$ be a combination of these
smooth functions, $f(x) =  \sum\limits_{k=1}^{\infty}e^{-\gamma_i x} f_i(x).$
Let $\left\{-\gamma_1,-\gamma_2  \right\}\in \mathbb{C}$ be the upper and lower critical exponents  of
 $f(x),$ that is, $ A_2 e^{ -x {\rm Re} \gamma_2 } < \left|f(x)\right| < A_1e^{-x {\rm Re} \gamma_1 }$ for $x \in ]-\infty,\infty[, $  where $A_1$ and $A_2$ are real constants. Also, let any function  $f_i(x)$ from this set of the smooth functions
be a function possessing a series representation with an infinite radius of convergence in the real domain. Finally, let the Laplace transform
\begin{eqnarray*}
L[f(x),x](z)  =  \int\limits_{0}^{\infty} f(x)e^{-zx} ~dx \no\\ 
\end{eqnarray*}
 be defined in the domain    $ {\rm Re}~(z  + \gamma_1) > 0$
 of the complex $z$ plane and may be continued analytically to the whole complex $z$ plane   $\mathbb{C}.$
 Then, the inverse integral transformation  
\begin{eqnarray*} 
f(x) = \frac{1}{2\pi i}\oint\limits_{C} e^{zx} L[f(x),x](z)~dz 
\end{eqnarray*}
recovers the function $f(x)$ for any  $x \in ]-\infty,\infty[.$ The contour $C$ is an arbitrary closed contour in the complex $z$ plane such that  all the poles  of  the transform  $L[f(x),x](z)$    in this  plane are inside the contour $C.$
\end{theorem}

\section{Mellin moments} \label{MMB}

We define Mellin $z$-moment of function $f(x)$ as  
\begin{eqnarray} \label{MMT} 
M[f(x),x](z) =  \int\limits_0^{1} x^{z-1}f(x)~dx,
\end{eqnarray}
$z$ is a complex variable.    To construct the inverse transformation, 
we need to rewrite (\ref{MMT}) in the form of the Laplace transform (\ref{LT}) and then to use  (\ref{LT Cauchy}), 
\begin{eqnarray*}  
L[f(x),x](z) = \int\limits_{0}^{\infty} e^{-xz}f(x)~dx =  \int\limits_{-\infty}^{0} e^{xz}f(-x)~dx \\
= \int\limits_{0}^{1} y^{z-1}f(-\ln y)~dy \no\\
\equiv \int\limits_{0}^{1} y^{z-1}F(y)~dy \equiv M[F(y),y](z),
\end{eqnarray*}
where we have introduced a new function $F(y) \equiv f(-\ln y).$  The Laplace transform for the function $f(x)$ appears to be
the Mellin moment for the function $F(y),$  
\begin{eqnarray} \label{MMT Cauchy}
 \dis{f(x) = \frac{1}{2\pi i}\int\limits_{a+\delta-i\infty}^{a+\delta+i\infty}e^{xz} L[f(x),x](z) ~dz}    \Rightarrow \dis{F(y)}  \no\\ 
= \dis{f(-\ln y) = \frac{1}{2\pi i}\int\limits_{a+\delta-i\infty}^{a+\delta+i\infty}y^{-z} L[f(x),x](z) ~dz} \no\\
 = \dis{\frac{1}{2\pi i}\int\limits_{a+\delta-i\infty}^{a+\delta+i\infty}y^{-z} M[F(y),y](z) ~dz }.
\end{eqnarray}

Since the Laplace transform  $L[f(x),x](z)$  is defined in the domain  ${\rm Re}~z > a,$ where $a$ is an index of the exponential growth of 
the function $f(x),$ the Mellin moment   $M[F(y),y](z)$ is defined in the same domain because the power-like restriction on its growth\footnote{In the case of the Mellin moments we call $a$ ``critical index''.}
\begin{eqnarray} \label{Power-like growth-MMT} 
\left|F(y)\right| < A/y^{a} 
\end{eqnarray}
comes from the restrictions on $f(x).$ In the inverse transformation   the contour passes vertically in the complex plane $z$
in the same position at ${\rm Re}~z = a + \delta$ as it does for the Laplace transformation (\ref{LT Cauchy}). Under this condition the  
$M[F(y),y](z)$ does not have poles in the complex half-plane to the right from this vertical line.  

The direct transformation proof may be done as 
\begin{eqnarray} \label{MO direct} 
M[F(y),y](z) \\
= \frac{1}{2\pi i}\int\limits_{0}^{1} y^{z-1} ~dy \int\limits_{a+\delta-i\infty}^{a+\delta+i\infty}y^{-u} M[F(y),y](u) ~du  \no \\
= \frac{1}{2\pi i}\int\limits_{a+\delta-i\infty}^{a+\delta+i\infty}\frac{M[F(y),y](u)}{z-u}du =  M[F(y),y](z), \no
\end{eqnarray}
and the inverse transformation may be proved as 
\begin{eqnarray} \label{MO inverse}
F(y) = \frac{1}{2\pi i}  \int\limits_{a+\delta-i\infty}^{a+\delta+i\infty}y^{-z}  M[F(y),y](z) dz \no\\
= \frac{1}{2\pi i}  \int\limits_{a+\delta-i\infty}^{a+\delta+i\infty}y^{-z} \int\limits_{0}^{1} u^{z-1} F(u) ~du dz  \no \\
= \int\limits_{0}^{1} \delta\le\ln{\frac{u}{y}}\ri F(u) u^{-1}~du  = F(y),
\end{eqnarray}
where $0 < y < 1.$  Thus, the transformation (\ref{MMT Cauchy}) is inverse to transformation (\ref{MMT}) 
under the restriction for the power-like growth  (\ref{Power-like growth-MMT}).

The Mellin moments, Laplace transforms and Mellin transforms posses the same equation for the inverse transformation. If all these three transforms may be calculated for the same function,
the inverse transformations would give 
the same result and the corresponding integrals would be related one to another by complex diffeomophisms in the complex planes of the variables of the inverse transformations.

\subsection{Inverse transformation of the moment $M[y^\gamma,y](z)$ in the domain $y\in[0,\infty[$ } \label{MMP}

For the future use, we need a Mellin $z$-moment of the power function $y^\gamma$ with respect to the variable $y.$  
This is a power-like function,  the Mellin transformation is impossible in this case, the integral  
\begin{eqnarray*}
\int\limits_0^{\infty}dy~y^{z-1} y^\gamma
\end{eqnarray*}
is divergent, however, the  Mellin moment with respect to $y$ is possible because we have a power-like restriction on the growth of the  functions,
\begin{eqnarray*} 
y^{{\rm Re}\gamma} < y^{{\rm Re}\gamma - \delta}, ~~~ \delta > 0, ~~~~ y \in[0,1]. 
\end{eqnarray*}
This power-like function   $y^\gamma$  may be used to show that the information about the values of function  $y^\gamma$ in an arbitrary positive domain of $y$ may be obtained 
from its Mellin $z$-moment $M[y^\gamma,y](z)$ defined in (\ref{MMT}) by applying an extended inverse transformation  to $M[y^\gamma,y](z)$  instead of the standard inverse
transformation of the Mellin $z$-moment  which is given by Eq.(\ref{MMT Cauchy}). The purpose of this Section \ref{MMB} is to construct such an extended  inverse transformation of the  Mellin $z$-moment.

Indeed, we consider first the case, when $y$ is in the standard domain $y \in [0,1].$  The Mellin $z$-moment  of   $y^\gamma$ with respect to  variable $y$ is 
\begin{eqnarray*} 
M[y^\gamma,y](z) =   \int\limits_0^1 y^\gamma y^{z-1}dy  = \frac{1}{\gamma + z}, 
\end{eqnarray*}
the domain of complex variable $z$ for taking this Mellin moment is ${\rm Re}(\gamma + z) > 0.$ In the complex plane of the variable 
$z$ the expression above may be analytically continued to all the complex plane $\mathbb{C}$ of  the variable $z.$  
The inverse operation (\ref{MMT Cauchy}) in this case is 
\begin{eqnarray} \label{y<1} 
y^\gamma = \frac{1}{2\pi i}\int\limits_{-{\rm Re} \gamma +\delta-i\infty}^{-{\rm Re} \gamma + \delta+i\infty}\frac{y^{-z}}{\gamma + z} ~dz,
\end{eqnarray} 
$\delta$ is a small positive real \footnote{We consider $\delta$ to be real infinitesimally small positive in all this paper.}. Because $y \in [0,1]$ we must
close the contour to the left complex infinity in order to avoid any contribution from this complex infinity and use the Cauchy formula due to which only one 
residue at $z=-\gamma$ contributes.

However, for future use in DGLAP equation  \cite{Alvarez:2016juq,Kondrashuk:2019cwi,Alvarez:2019eaa,Kondrashuk:2025ajb} we need to work with the extended domain $y \in [0,\infty[$ because one of two variables of this IDE  runs in this
domain (see Section \ref{Sec-pro}).
This DGLAP variable is called momentum transfer. We may modify the contour to the form  
\begin{eqnarray} \label{y>1} 
y^\gamma = - \frac{1}{2\pi i} \int\limits_{-{\rm Re} \gamma - \delta-i\infty}^{-{\rm Re} \gamma - \delta+i\infty}\frac{y^{-z}}{\gamma + z} ~dz,
\end{eqnarray} 
for the domain $y \in [1,\infty[.$ The integral with this contour is equal to zero for  the domain $y \in [0,1]$ because in such a case 
we should close it to the left complex infinity in order to avoid the  contribution of the complex infinity,  however in such a case we do not have any contribution 
of the residues inside the contour. In the case when $y \in [1,\infty[,$ we need to close the contour to the right infinity in order to avoid 
the contribution of the complex infinity and in this case the only residue that contributes is  $z=-\gamma.$

\begin{figure}[ht!] 
\centering\includegraphics[scale=0.4]{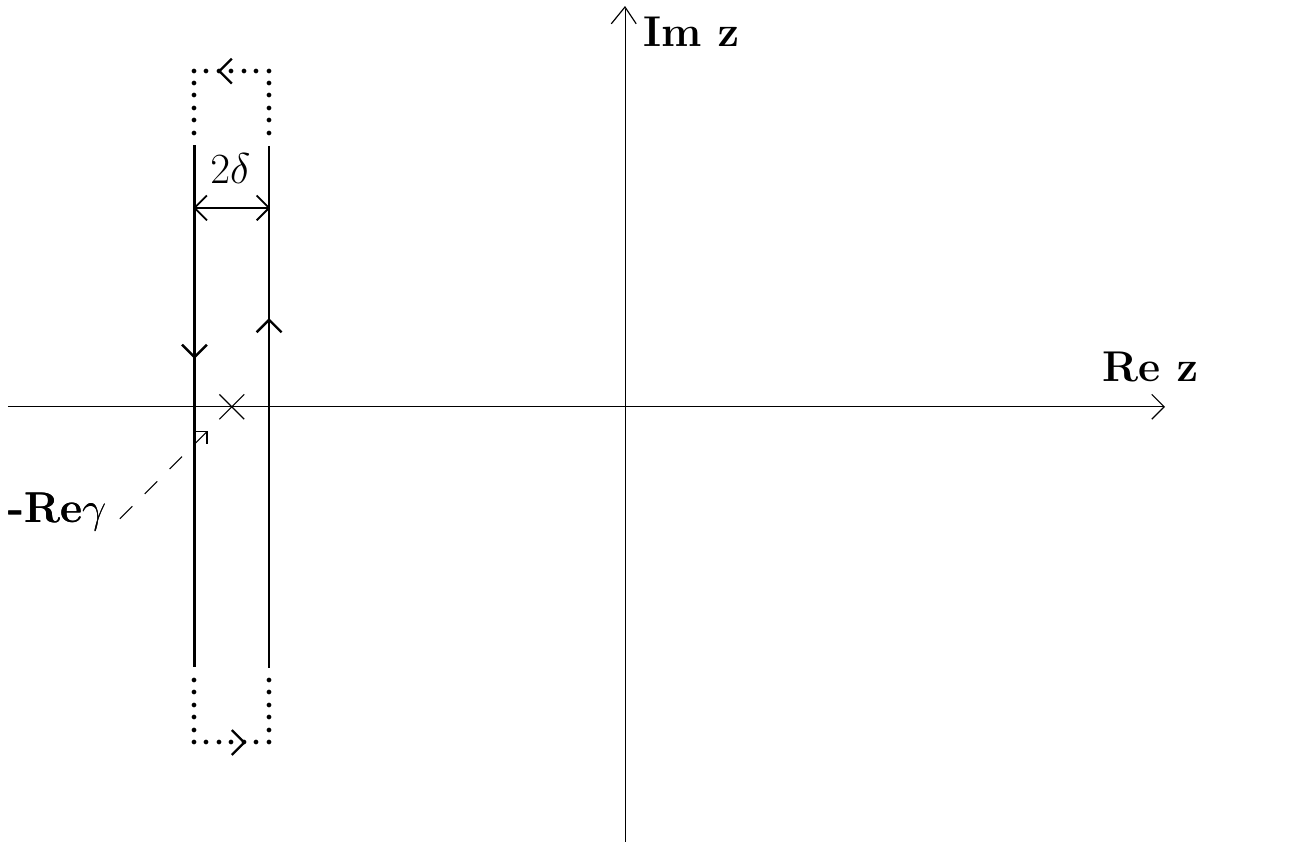}
\caption{\footnotesize  Contour $CR$  for the Laplace transform $M[y^{\gamma}, y](z)$}
\label{FMMP}
\end{figure}

To conclude, integral (\ref{y<1}) gives  $y^\gamma$  for the domain $y \in [0,1]$ and gives zero for the domain $y \in [1,\infty[,$ while
integral (\ref{y>1}) gives  $y^\gamma$ for the domain $y \in [1,\infty[$ and gives zero for the domain $y \in [0,1].$ In the second case 
the negative sign disappears due to clockwise orientation of the contour. Thus, we may write 
\begin{eqnarray}  \label{y>0} 
y^\gamma =  \frac{1}{2\pi i}\int\limits_{-{\rm Re} \gamma +\delta-i\infty}^{-{\rm Re} \gamma + \delta+i\infty}\frac{y^{-z}}{\gamma + z} ~dz  \no\\
-  \frac{1}{2\pi i}\int\limits_{-{\rm Re} \gamma - \delta-i\infty}^{-{\rm Re} \gamma - \delta+i\infty}\frac{y^{-z}}{\gamma + z} ~dz 
\end{eqnarray}
and taking into account that $\delta \rar 0$ is infinitesimally small to avoid the contribution of the smallest sides of the rectangular 
shown in Fig. \ref{FMMP}, we may re-write this equality  as 
\begin{eqnarray} \label{MCR} 
y^{\gamma} = \frac{1}{2\pi i}\oint\limits_{CR} \frac{y^{-z}}{\gamma + z} ~dz,  ~~~~ y \in[0,\infty[ 
\end{eqnarray}
where we have a rectangular contour $CR$ which contains two straight vertical lines, one line crosses the real axis at the point at $z= -{\rm Re} \gamma +\delta$ and 
another crosses the real axis at the point $z = -{\rm Re} \gamma - \delta.$ In the right line the integration is performed from the negative imaginary 
infinity to the positive imaginary infinity, while for the left line we integrate down from the positive imaginary infinity to 
the negative imaginary infinity. This integration corresponds exactly to the counterclockwise orientation of the contour and by Cauchy formula 
corresponds to the only residue at the point $z= -\gamma$ in the complex plane of the variable $z.$  
Strictly speaking,  it is not necessary to make the size of 
the smallest sides of the rectangular contour to be infinitesimally small in order to construct the extended  contour (\ref{MCR}) of the inverse Mellin $z$-moment. 
In Subsection \ref{LTE} we have mentioned that this requirement may be discarded when we applied the same contour modification in order to  extend the inverse 
Laplace transformation of the Laplace transform  $L[e^{-\gamma x},x](z).$ In Subsection \ref{LTG} we  have already considered the rectangular contour with the finite size of the smallest sides. 
Eq.(\ref{MCR}) is valid for any positive real $\delta$ and  any real positive $y,$  that is, $y \in~ [0,\infty[.$ 
The extended inverse transformation  (\ref{MCR}) recovers the power $y^{\gamma}$ for any $y$ in the extended domain  $[0,\infty[.$

We may repeat the direct transformation proof (\ref{MO direct}) we have found for the standard domain $y \in[0,1]$ from the definition  (\ref{MMT}) and apply it for the extended domain $y \in[0,\infty[,$    
\begin{eqnarray}  \label{MCR direct} 
\frac{1}{z+\gamma} = \int\limits_{0}^{1} y^{\gamma}y^{z-1} ~dy \no\\
= \frac{1}{2\pi i}\int\limits_{0}^{1}y^{z-1}~dy\oint\limits_{CR} \frac{y^{-\omega}}{\gamma + \omega} ~d\omega \no\\
= \frac{1}{2\pi i}\oint\limits_{CR} \frac{1}{(\gamma + \omega)(z-\omega)} ~d\omega = \frac{1}{z+\gamma}.
\end{eqnarray}
Here the calculation of the residues may be done inside or outside the contour, the result will be the same.  This may be proven that one of these two ways of calculation is equivalent to another. It is supposed
in Eq. (\ref{MCR direct}) that we are in the domain ${\rm Re}(\gamma + z) > 0$ of the complex plane of the Mellin moment $z.$

Also, we may repeat the inverse transformation proof   (\ref{MO inverse})  which we have found for the standard domain $y \in[0,1]$  and apply it for the extended domain $y \in[0,\infty[,$ 
\begin{eqnarray}  \label{MCR inverse} 
y^{\gamma} =  \frac{1}{2\pi i}\oint\limits_{CR} \frac{y^{-z}}{\gamma + z} ~dz  \\
= \frac{1}{2\pi i} \left[\int\limits_{-{\rm Re} \gamma +\delta-i\infty}^{-{\rm Re} \gamma + \delta+i\infty}y^{-z}~dz\int\limits_0^1  u^{\gamma + z-1} ~du \right. \no\\
%\end{eqnarray}
%\begin{eqnarray*}
\left. 
+ \int\limits_{-{\rm Re} \gamma - \delta-i\infty}^{-{\rm Re} \gamma - \delta+i\infty}y^{-z}~dz\int\limits_1^{\infty}  u^{\gamma + z-1} ~du \right]   \no\\
= \int\limits_0^1 \delta\le\ln{\frac{u}{y}}\ri u^{\gamma -1} ~du + \int\limits_1^{\infty} \delta\le\ln{\frac{u}{y}}\ri  u^{\gamma -1} ~du   \no\\
 =  \int\limits_0^{\infty} \delta\le\ln{\frac{u}{y}}\ri  u^{\gamma -1} ~du. \no
\end{eqnarray}
This chain of equalities is valid for any real  positive $y.$ 

Thus, in this Subsection we have generalized the standard inverse  transformation (\ref{y<1}) of the  Mellin moment  $M[y^\gamma,y](z)$ from the standard domain 
$y \in[0,1]$  to the extended inverse transformation  (\ref{MCR}) of the Mellin $z$-moment  $M[y^\gamma,y](z)$ which reproduces the power   $y^{\gamma}$   for any $y$ from the extended domain  $y \in[0,\infty[.$

\subsection{Inverse transformation of the moment $M[F(y),y](z)$ in the domain $y\in[0,\infty[$ }  \label{MMG}

In Subsection \ref{MMP} we considered a power-like function $F(y) = y^{\gamma},$ took its Mellin moment and modified the inverse transformation of the Mellin $z$-moment in such a way that 
it became possible to recover the original function  $y^{\gamma}$ in all  the range of real positive $y$ by this extended inverse transformation of the Mellin $z$-moment $M[y^\gamma,y](z).$
We should mention that the standard inverse transformation of the Mellin $z$-moment (\ref{MO inverse})
may recover the original function  $y^{\gamma}$ only for the standard domain $y \in~[0,1].$  Such a generalization for the power-like function appears to be 
important for the future use in quantum chromodynamics when we will apply this extended inverse transformation  of $M[y^\gamma,y](z)$ to construct an IDE dual 
to the DGLAP equation  \cite{Alvarez:2016juq,Kondrashuk:2019cwi,Alvarez:2019eaa,Kondrashuk:2025ajb} (see Section \ref{Sec-pro}).
However, it would be helpful to write an analog of Eq.(\ref{MCR}) to reproduce an arbitrary function $F(y)$ in the extended domain $y \in[0,\infty[,$ too,  
after making the extended inverse transformation   of the Mellin $z$-moment  $M[F(y),y](z)$  that is defined in Eq. (\ref{MMT Cauchy}).
In the rest of this Subsection we prove a possibility to modify the contour of the inverse transformation of the Mellin $z$-moment  $M[F(y),y](z)$ of an arbitrary  function  $F(y)$ in order to reach this purpose.

\begin{figure}[hbt!!!] 
\centering\includegraphics[scale=0.4]{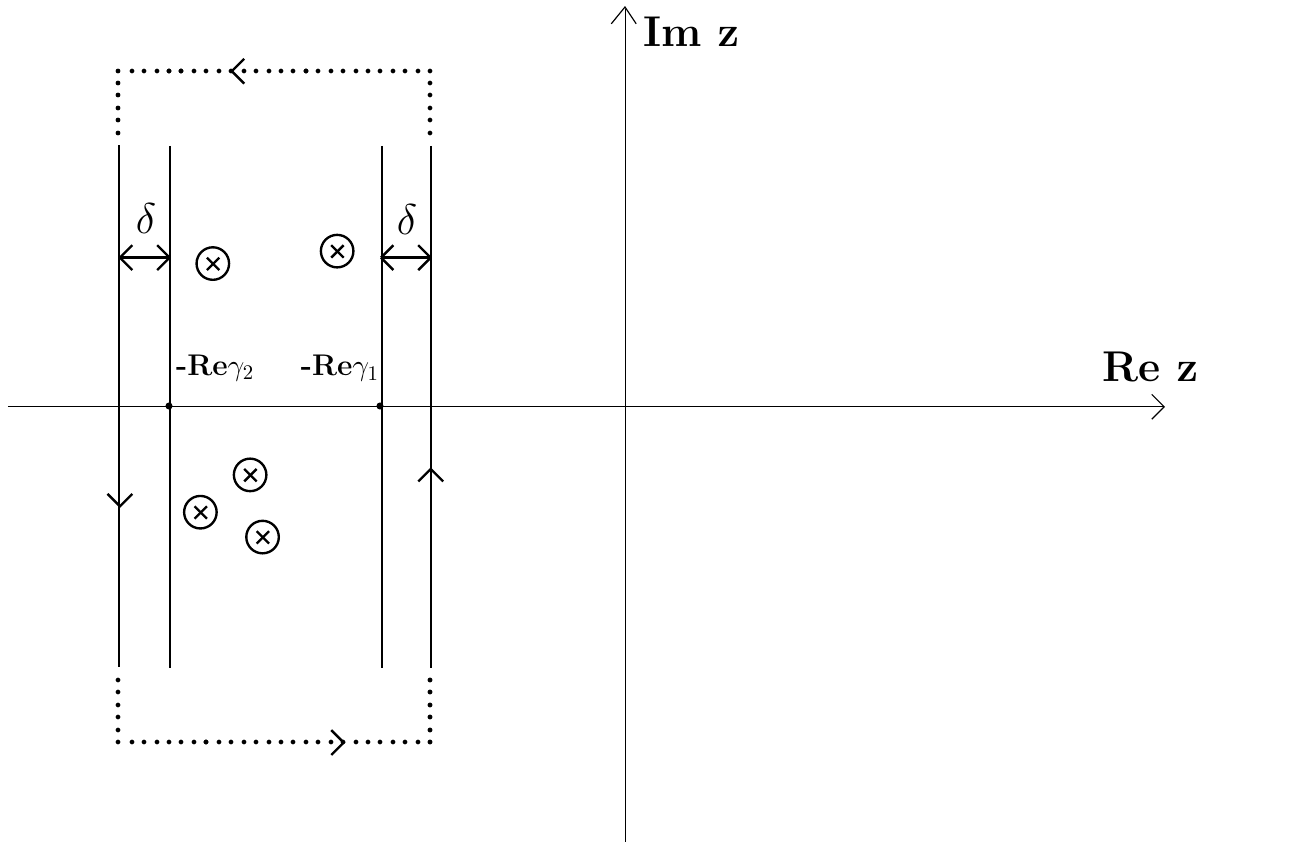}
\caption{\footnotesize  Contour $CR$ for the Mellin $z$-moment  $M[F(y),y](z)$ with poles inside }
\label{FMMG}
\end{figure}

First, let us start with the standard inverse transformation (\ref{MMT Cauchy}) for the standard domain $y \in ~[0,1],$ and that $-{\rm Re}\gamma_1$ is the critical index of the function 
$F(y)$ from Eq.(\ref{Power-like growth-MMT}),  $M[F(y),y](z)$ is defined for ${\rm Re}~z  > -{\rm Re} \gamma_1.$ This means that all the poles are situated to the left with respect to 
the critical index in the complex plane of the variable $z.$ We continue analytically  $M[F(y),y](z)$  to the whole complex plane $z \in \mathbb{C}$
and suppose that number of poles to the left from the vertical line of the transformation (\ref{MMT Cauchy}) is countable. 
This means that we may draw the second vertical line which passes a bit to the left of the leftmost pole in the complex plane $z,$  
and we get a rectangular contour  drawn in Fig.\ref{FMMG}.
All the poles of the Mellin moment $M[F(y),y](z)$ in Eq. (\ref{MMT Cauchy}) are inside  this contour in the complex plane $z.$ 
The second vertical line crosses the real axis of the plane $z$
at the point $-{\rm Re}~\gamma_2$ in Fig. \ref{FMMG}. This contour is closed to the rectangular form by two horizontal lines at the imaginary complex infinities in the strip  
$ - {\rm Re}~\gamma_2 < {\rm Re}~z  < -{\rm Re} \gamma_1.$

If $y<1$ we need to close each contour associated to every of these two vertical lines to the left complex infinity for both the vertical lines in order to avoid the contribution of this complex infinity, 
in such a case the right vertical line contributes with all the residues 
on the left hand side of it, and the left vertical line does not contribute at all because there is no residue on the left hand side of it by construction of this contour.  This analysis repeats exactly 
Eq. (\ref{y<1}) which we have written in Subsection \ref{MMP} dedicated to the power-like function. 
If $y>1$ we need to close each contour associated to every of these two vertical lines to the right complex infinity for both the vertical lines, in such a case the left vertical line contributes with all the residues 
on the right hand side of it, and the right vertical line does not contribute at all because there is no residue on the right hand side of it by construction of this contour.  This consideration is given in complete analogy
to Eq. (\ref{y>1}) which we have written in Subsection \ref{MMP} dedicated to the power-like function.  Whether both the contours associated to these vertical lines are closed to the right complex infinity or 
they are closed to the left complex infinity, the contribution of the residues will be the same and the result of this residue calculus will be $F(y).$  All this is written in complete analogy 
to the case of power-like function considered in Subsection \ref{MMP} in Eq. (\ref{y>0}).

Thus, we may write 
\begin{eqnarray*}   
F(y)  = \frac{1}{2\pi i}\int\limits_{-{\rm Re} \gamma_1 +\delta-i\infty}^{-{\rm Re} \gamma_1 + \delta+i\infty} y^{-z} M[F(y),y](z)  ~dz \no\\
- \frac{1}{2\pi i}\int\limits_{-{\rm Re} \gamma_2 - \delta-i\infty}^{-{\rm Re} \gamma_2 - \delta+i\infty} y^{-z} M[F(y),y](z) ~dz 
\end{eqnarray*}
and then may close the contour  to the rectangular form at the complex imaginary infinities of the strip $-{\rm Re} \gamma_2 - \delta <  {\rm Re}~z  < -{\rm Re} \gamma_1 + \delta$  as it is depicted in Fig. \ref{FMMG}.
As the result, we obtain complete analog of formula (\ref{MCR}) in which instead of the Mellin $z$-moment $M[y^\gamma,y](z) = 1/(\gamma + z)$ the Mellin $z$-moment $M[F(y),y](z)$  of an arbitrary function $F(y)$ 
is written,
\begin{eqnarray} \label{MCRG} 
F(y) = \frac{1}{2\pi i}\oint\limits_{CR} y^{-z} M[F(y),y](z)~dz,  ~~~~ y \in[0,\infty[. 
\end{eqnarray}
Here it is worthy to mention that the Mellin moment  $M[F(y),y](z)$  is  always defined by Eq. (\ref{MMT}). The rectangular contour $CR$  is depicted in Fig. \ref{FMMG}.
This contour contains two vertical lines, one line crosses the real axis at the point at $z= -{\rm Re} \gamma_1 +\delta$ and 
another line crosses the real axis at the point $z = -{\rm Re} \gamma_2 - \delta.$ In the right line the integration is performed from the negative imaginary 
infinity to the positive imaginary infinity, while for the left line we integrate down from the positive imaginary infinity to 
the negative imaginary infinity. This integration corresponds exactly to the counterclockwise orientation of the contour and by Cauchy integral formula 
corresponds to the contribution of all the residues of   $M[F(y),y](z)$ in the strip  $-{\rm Re} \gamma_2 - \delta <  {\rm Re}~z  < -{\rm Re} \gamma_1 + \delta$   
in the complex plane of the variable $z.$   Eq.(\ref{MCRG}) is valid for any positive real $\delta$ and $y,$  that is, $y \in[0,\infty[.$ 
The extended inverse transformation  (\ref{MCRG}) of  $M[F(y),y](z)$ recovers the function $F(y)$ in the extended domain  $y \in [0,\infty[.$

We may repeat the direct transformation proof (\ref{MCR direct}) of Subsection \ref{MMP} which we have found for the power-like function in the extended domain $y \in[0,\infty[,$
\begin{eqnarray} \label{MCRG direct} 
M[F(y),y](z)  = \int\limits_{0}^{1} F(y)y^{z-1} ~dy \no\\
= \frac{1}{2\pi i}\int\limits_{0}^{1}y^{z-1}~dy\oint\limits_{CR} y^{-\omega} M[F(y),y](\omega)  ~d\omega \no\\
= \frac{1}{2\pi i}\oint\limits_{CR} \frac{M[F(y),y](\omega)}{z-\omega} ~d\omega = M[F(y),y](z).
\end{eqnarray}
Here the calculation of the residues may be done inside or outside the contour. It may be proven that residue calculus inside the contour and the residue calculus outside the contour give the same results. 
This  formulae suppose that we are in the standard domain ${\rm Re}(\gamma_1 + z) > 0$ of the complex plane of the Mellin moment $z.$

Also, we may repeat the inverse transformation  proof  (\ref{MCR inverse})  which we have found in Subsection \ref{MMP} for the extended  domain $y \in[0,\infty[$ of the inverse transformation of the Mellin 
$z$-moment $M[y^\gamma,y](z)$ which was based on the formula (\ref{y>0}),
\begin{eqnarray}  \label{MCRG inverse}
F(y)  =  \frac{1}{2\pi i}\oint_{CR} y^{-z} M[F(y),y](z)~dz  \no\\
= \frac{1}{2\pi i} \left[\int\limits_{-{\rm Re} \gamma_1 +\delta-i\infty}^{-{\rm Re} \gamma_1 + \delta+i\infty}y^{-z}~M[F(y),y](z)~dz \right.\no\\
+ \left. \int\limits_{-{\rm Re} \gamma_2 - \delta+i\infty}^{-{\rm Re} \gamma_2 - \delta-i\infty}y^{-z}~M[F(y),y](z)~dz \right] \no\\
= \frac{1}{2\pi i} \left[\int\limits_{-{\rm Re} \gamma_1 +\delta-i\infty}^{-{\rm Re} \gamma_1 + \delta+i\infty}y^{-z}~dz\int\limits_0^1 F(u) u^{z-1} ~du \right.\no\\ 
-  \left.\int\limits_{-{\rm Re} \gamma_2 - \delta-i\infty}^{-{\rm Re} \gamma_2 - \delta+i\infty}y^{-z}~dz\int\limits_1^{\infty} F(u) u^{z-1} ~du \right]  \no\\
= \int\limits_0^1 \delta\le\ln{\frac{u}{y}}\ri F(u) u^{ -1} ~du 
+ \int\limits_1^{\infty} \delta\le\ln{\frac{u}{y}}\ri  F(u)u^{ -1} ~du   \no\\
= \int\limits_0^{\infty} \delta\le\ln{\frac{u}{y}}\ri F(u) u^{ -1} ~du. 
\end{eqnarray}
This chain of equalities is valid for any real  positive $y.$ The replacement of  $M[F(y),y](z)$ defined in (\ref{MMT}) with a bit different expression $-\int_1^{\infty} F(u) u^{z-1} ~du$ in the second
integral of (\ref{MCRG inverse}) is justified by the residue calculus.  This would be just a generalization of the case   $M[y^{\gamma}, y](z)$ considered in Eq. (\ref{MCR inverse}) 
to the Mellin moment of an arbitrary function $M[F(y), y](z).$   Here  $-\gamma_1$ is the rightmost pole in the complex plane $z$ and $-\gamma_2$ is the leftmost pole 
in the complex plane of $z$-moment  $M[F(y),y](z).$ This means that the upper critical index is $-{\rm Re} \gamma_1 + \delta$ and the lower critical index $-{\rm Re} \gamma_2 - \delta.$
For the case when $y \in[0,1]$  that is, the variable $y$ is in the standard domain of the transformation (\ref{MMT}) we may reproduce the inverse transformation proof  (\ref{MO inverse}) 
from this proof (\ref{MCRG inverse}) of the extended inverse  transformation of   the Mellin $z$-moment.

\subsection{Summarizing Mellin moments} \label{smz moment}

Finally, at the end of this Subsection we would like to do three summarizing comments.  

\begin{itemize}
\item The domain of variable $y$ of  $F(y)$ should include the interval $y \in [0, 1],$ otherwise the Mellin $z$-moment transformation  (\ref{MMT})
would be impossible to define. In brief, summarizing the discussion,  the transformation of the function $F(y)$ to the Mellin $z$-moment  $M[F(y),y](z)$
is defined in the domain  ${\rm Re}~z > a,$ where $a$ is an index of the power-like growth of the function $F(y)$ 
\begin{eqnarray*}
\left|F(y)\right| < A/y^{a}. 
\end{eqnarray*}
In the standard inverse transformation (\ref{MMT Cauchy}) from the Mellin $z$-moment  $M[F(y),y](z)$ to the function $F(y)$ the contour passed vertically in the complex plane $z$ at ${\rm Re}~z = a + \delta.$ 
Under this condition the  moment $M[F(y),y](z)$ does not have poles in the complex half-plane to the right from this vertical line  in the complex plane of the variable $z.$  
\item A power-like upper bound for the dependence on the variable $y$ is the necessary condition for taking the Mellin moment of $F(y).$ 
In case if the lower bound for the power-like behaviour exist, the contour in the complex plane $z$ of the function  $M[F(y),y](z)$ would contain two vertical lines 
in such a manner that the left one is a bit to the left from the lower bound value, the right one is a bit to the right from the 
upper bound value on the real axis of the complex   plane of the variable $z.$  The contour of this type is shown in  Fig. (\ref{FMMG}).
An example of such a type of the functions would be $y^{\gamma_1} + y^{\gamma_2}.$ In a general case,
we suppose that the number of residues inside the contour is countable. The positions of the vertical lines of the contour depend on the bounds of the function $F(y)$ with respect 
to the variable $y,$ it may even contain the left complex infinity.
The function $M[F(y),y](z)$  may be continued analytically from the domain in which it is defined  
to all the complex plane of the variable $z.$  We may use this analytic continuation in order to recover the information for $F(y)$ for an arbitrary real positive domain of the 
variable $y$  from its Mellin $z$-moment  $M[F(y),y](z)$  (\ref{MMT}).
\item To determine how the poles in the complex plane of variable $z$ are distributed, we need more information
about the function $F(y).$   
We have obtained the extended inverse transformation  of the Mellin $z$-moment  $M[F(y),y](z)$  in Subsection \ref{MMG} in which the contour of this inverse transformation has a rectangular form. We may change 
the form of the border of this rectangular contour in any way under the condition that all the poles remain inside it. Then by Cauchy formula the result will be the same.
We fix the final form of the dual contour  from the considerations based on the DGLAP equation 
where the contour $C$ will have a form different from the rectangular  \cite{Kondrashuk:2019cwi,Alvarez:2019eaa,Kondrashuk:2025ajb}, 
\begin{eqnarray*} 
F(y) =  \frac{1}{2\pi i}\oint\limits_{C}~dz y^{-z} M[F(y),y](z)
\end{eqnarray*}
but with all the residues inside the contour. 

\end{itemize}

\subsection{Two theorems proven in Section \ref{MMB}}

\begin{theorem}\label{Moment-1}
Let $\gamma \in \mathbb{C}$  and the formula 
\begin{eqnarray*} 
M[y^\gamma,y](z) =   \int\limits_0^1 y^\gamma y^{z-1}dy  = \frac{1}{\gamma + z}
\end{eqnarray*}
be the Mellin moment  of the function $y^{\gamma}$ defined in the complex plane of the variable $z$  for  ${\rm Re}(\gamma + z) > 0.$  Then, the
inverse integral transformation of the Mellin moment is  
\begin{eqnarray*} 
y^{\gamma} = \frac{1}{2\pi i}\oint\limits_{C} \frac{y^{-z}}{\gamma + z} ~dz,  ~~~~ y \in[0,\infty[ 
\end{eqnarray*}
recovers the function $y^{\gamma}$ for any  $y \in [0,\infty[.$ The contour $C$ is an arbitrary closed contour in the complex  $z$ plane such that  the point $z = -\gamma$ is inside the contour.
\end{theorem}

\begin{theorem}\label{Moment-2}
Let functions $F_i(y)$ be an infinite or finite  set of smooth functions  of the real argument,
$F_i(y)\in C^{\infty}\le  [0,\infty[ \ri.$ Let a function $F(y)$ be a combination of these
smooth functions, $F(y) =  \sum\limits_{k=1}^{\infty}y^{\gamma_i} F_i(y).$
Let $\left\{-\gamma_1,-\gamma_2  \right\}\in \mathbb{C}$ be the upper and lower critical indices  of
 $F(y),$ that is, $ A_2y^{ {\rm Re} \gamma_2 } < \left|F(y)\right| < A_1y^{ {\rm Re} \gamma_1 }$ for $y \in [0,1], $  where $A_1$ and $A_2$ are real constants. Also, let any function  $F_i(y)$ from this set of the smooth functions
be a function possessing a series representation with an infinite radius of convergence in the real domain.  Finally, let the  Mellin $z$-moment
\begin{eqnarray*} 
M[F(y),y](z) =  \int\limits_0^{1} y^{z-1}F(y)~dy
\end{eqnarray*}
be defined in the domain    $ {\rm Re}~(z  + \gamma_1) > 0$
 of the complex $z$ plane and may be continued analytically to the whole complex $z$ plane   $\mathbb{C}.$
Then, the inverse integral transformation  
\begin{eqnarray*} 
F(y) = \frac{1}{2\pi i}\oint\limits_{C} y^{-z} M[F(y),y](z)~dz 
\end{eqnarray*}
recovers the function $F(y)$ for any  $y \in [0,\infty[.$ The contour $C$ is an arbitrary closed contour in the complex $z$ plane such that  all the poles  of  the Mellin moment  $M[F(y),y](z)$    in this  plane are inside the contour $C.$
\end{theorem}

\section{Examples} \label{examples}

In this Section we consider  examples which show  explicitly that Theorem \ref{Moment-2} is valid. The
 cases of a finite number of poles in the complex plane of the inverse transformation variable $z$ and  of the infinite radius of convergence for the series representation of the transforming function $F(y)$ around  the point $y=0$ are considered first.    Another approach is necessary for the cases when the series representation  of the function  $F(y)$ in the domain $y \in [0,1]$ does not converge when we apply  it for   $y>1.$  In such cases we have to modify the integrands of the inverse transformations for the Mellin moments
in order to include both the cases $y \in [0,1]$ and $y \in [1,\infty[$ on the same foot. 

\subsection{Polynomial} \label{poly}

The simplest example would be a binomial, for example $F(y) = (1-y)^2.$ Its $z$-moment is  Euler beta function,     
\begin{eqnarray*} 
\int\limits_{0}^{1} y^{z-1}(1-y)^2 dy = \frac{\G(z)\G(3)}{\G(z+3)} = \frac{2}{(z+2)(z+1)z}. 
\end{eqnarray*}
The inverse transformation from Theorem \ref{Moment-2} calculates  three residues at the poles $z=0, z=-1, z=-2$  and recovers the original function  back
\begin{eqnarray} \label{value} 
F(y) = \frac{1}{2\pi i}\oint\limits_{C} \frac{2y^{-z}dz}{(z+2)(z+1)z}  = 1 -2y +y^2. 
\end{eqnarray}
According to our construction, the contour $C$ encloses all the three residues. The value of $y$ in Eq.  (\ref{value}) does not play any role
because we have here a finite number of residues (three residues in this case). The result   (\ref{value})  is valid for any  $y \in [0,\infty[.$

\subsection{A series with an infinite radius of convergence }  \label{infinite}

Here is an example with the exponential function. It is well-known that this function has a series representation with an infinite radius of convergence, 
\begin{eqnarray} \label{infi} 
\int\limits_{0}^{1} y^{z-1}e^{y}~ dy = \sum_{k=0}^{\infty}  \int\limits_{0}^{1} \frac{y^{z+k-1}}{k!}~ dy 
=  \sum_{k=0}^{\infty}  \frac{1}{k!(z+k)}.  
\end{eqnarray}
The inverse transformation is simple and recovers the function $e^y$  for any value of $y \in [0,\infty[,$
\begin{eqnarray} \label{series}
 \frac{1}{2i\pi}  \int\limits_{\delta-i\infty}^{\delta + i\infty} y^{-z}\sum_{k=0}^{\infty}  \frac{dz}{k!(z+k)}  
= \frac{1}{2i\pi} \sum_{k=0}^{\infty} \int\limits_{\delta-i\infty}^{\delta + i\infty}\frac{y^{-z}dz}{k!(z+k)} \no\\
 =  \sum_{k=0}^{\infty} \frac{y^{k}}{k!} = e^y.
\end{eqnarray}
The moment   (\ref{infi})  has infinitely many poles on the left half-plane,  however the resulting series  in the inverse transformation  (\ref{series}) is convergent for any  $y \in [0,\infty[.$ The standard inverse integral transformation for the moment (\ref{MMT Cauchy})
is to be  zero for $y>1$ by definition because there is no singularity on the right complex half-plane. The proposed in Theorem \ref{Moment-2}
modification of the inverse transformation recovers the value of the function $F(y)= e^y$ for any  $y \in [0,\infty[$ because all the poles in the complex $z$ plane remain inside the contour,
   \begin{eqnarray*}
 \frac{1}{2i\pi}  \oint\limits_C y^{-z}\sum_{k=0}^{\infty} \frac{dz}{k!(z+k)}  
=  \frac{1}{2i\pi} \sum_{k=0}^{\infty} \oint\limits_C\frac{y^{-z}dz}{k!(z+k)} \\
 =  \sum_{k=0}^{\infty} \frac{y^{k}}{k!} = e^y.
\end{eqnarray*}
It is just the Cauchy integral formula in this case of an infinite radius of convergence of the transforming function $e^y$ around any  point in its domain.

\subsection{Comparing  $z$-moments  with Mellin-Barnes transforms }  \label{comparing}

In this Subsection we consider Euler beta function with infinitely many poles on the left hand side from the imaginary axis is treated as
a Mellin moment. We have shown in Subsection \ref{poly} that  Theorem \ref{Moment-2} is valid for a Mellin moment with a finite number of poles in the complex $z$-plane of the transformation variable.  In this case it is the Cauchy integral formula. Now we consider a similar example of the Euler beta function but with an infinite number of poles in the complex $z$ plane,
\begin{eqnarray} \label{beta-2}
\int\limits_{0}^{1} y^{z-1}(1-y)^\frac{2}{3} dy = \frac{\G(z)\G\le\frac{5}{3}\ri}{\G\le z+ \frac{5}{3} \ri}. 
\end{eqnarray}
The function $F(y)  =  (1-y)^\frac{2}{3}  \notin C^{1}(  [0,\infty[ ),$ its derivative is infinite at $ y =1$ and the conditions of Theorem \ref{Moment-2}  are not fulfilled.  The discontinuity of the derivative at some points in the domain $[0,\infty[$ of the transformed function $F(y)$
means a finite radius of convergence for its series representation.

The series representation for the transforming functions $F(y)$ for most of the functions  suggests an infinite number of poles in the complex $z$ plane of its Mellin moment. In subsection \ref{smz moment}  we have constructed a rectangular contour and mentioned  that the left straight vertical line in the complex $z$ plane in the case of absence the leftmost pole passes on the left complex infinity. All the arguments presented in  Subsection \ref{MMG} perfectly work in that case when the leftmost pole is absent, too. However, we have supposed an infinite radius of  convergence of the resulting series in terms  of powers of $y$ obtained from the residues in the complex  $z$ plane. This is not the case of the transforming function $F(y)$  in Eq.  (\ref{beta-2}).

The standard definition  (\ref{MMT}) of the Mellin moments does not require an infinite radius of convergence of the transforming function $F(y).$ In the domain $y \in [0,1]$ the standard  inverse integral transformation  (\ref{MMT Cauchy})  of the Mellin moment can be applied and we obtain the original function back,
\begin{eqnarray}\label{moment-barnes-1}
 \frac{\G\le\frac{5}{3}\ri}{2i\pi}  \int\limits_{\delta-i\infty}^{\delta + i\infty} \frac{y^{-z}\G(z)}{\G\le z+ \frac{5}{3} \ri}  ~dz  \\
 =  \sum_{k=0}^{\infty}\frac{\Gamma\le-\frac{2}{3}+k\ri}{\Gamma\le-\frac{2}{3}\ri}\frac{y^k}{k!}  = (1-y)^\frac{2}{3}. \no
\end{eqnarray}

The contour should be closed to the left complex infinity because we suppose  in (\ref{moment-barnes-1})
the standard domain $0<y<1$ for the Mellin moments. Then we calculate the residues inside the contour.   The first two residues are
\begin{eqnarray*}
\res_{z= 0} \frac{y^{-z}\G(z)}{\G\le z+ \frac{5}{3} \ri}  =    \frac{1}{\G\le\frac{5}{3} \ri}, ~~~ 
\res_{z= -1} \frac{y^{-z}\G(z)}{\G\le z+ \frac{5}{3} \ri}  =    -\frac{y}{\G\le\frac{2}{3} \ri}.
\end{eqnarray*}

On the right hand side of (\ref{moment-barnes-1}) we use the well-known binomial expansion \cite{Alvarez:2016juq} for an arbitrary complex power $\gamma$ and $y \in [0,1],$
 \begin{eqnarray}  \label{binom}
(1-y)^\gamma =  \sum_{k=0}^{\infty}\frac{\Gamma(-\gamma+k)}{\Gamma(-\gamma)}\frac{y^k}{k!} = \sum_{k=0}^{\infty} \frac{(-\gamma)_k}{k!} y^k, 
\end{eqnarray}
in which Pochhammer symbol $(a)_k = \Gamma(a+k)/\Gamma(a)$ is involved. This formula may be derived by using Mellin-Barnes (MB) transformation \cite{Diaz:2024tdp}
\begin{eqnarray} \label{MBG}
\frac{1}{(1+y)^{\gamma}} = \frac{1}{2i\pi}  \frac{1}{\G(\gamma)} \int\limits_{\delta-i\infty}^{\delta + i\infty}y^{-z} \G(z)\G(\gamma -z)~dz \no\\
   = \frac{1}{2i\pi}  \frac{1}{\G(\gamma)} \int\limits_{-\delta-i\infty}^{-\delta + i\infty}y^{z} \G(-z)\G(\gamma +z)~dz   
\end{eqnarray}
by calculating the residues. This representation is valid for any real $y \in [0,\infty[$ and for any complex $\gamma$ with a small modification of the contour if necessary \cite{Diaz:2024tdp}. In particular,
for the function  $F(y)  =  (1-y)^\frac{2}{3}$ the MB representation
\begin{eqnarray} \label{moment-barnes-2}
(1-y)^{\frac{2}{3} } \no\\
= \frac{1}{2i\pi}  \frac{1}{\G\le-\frac{2}{3} \ri} \int\limits_{\delta-i\infty}^{\delta + i\infty}(-y)^{-z} \G(z)\G\le-\frac{2}{3} -z\ri dz 
\end{eqnarray}
can  be written. For $y \in [0,1]$ we close the integration contour to left complex infinity while for   $y \in [1,\infty[$ we close the integration contour to the right complex infinity just to avoid any contribution from the complex infinities \cite{Allendes:2009bd}. By comparing (\ref {moment-barnes-1})  and  (\ref {moment-barnes-2})  we conclude that for $y \in [0,1]$
\begin{eqnarray}  \label{vabene}
\frac{1}{2i\pi}  \frac{1}{\G\le-\frac{2}{3} \ri} \int\limits_{\delta-i\infty}^{\delta + i\infty}(-y)^{-z} \G(z)\G\le-\frac{2}{3} -z\ri dz  \\
 = (1-y)^{\frac{2}{3} }  = \frac{\G\le\frac{5}{3}\ri}{2i\pi}  \int\limits_{\delta-i\infty}^{\delta + i\infty} \frac{y^{-z}\G(z)}{\G\le z+ \frac{5}{3} \ri}  ~dz. \no
\end{eqnarray}
This identity is not valid for  $y \in [1,\infty[$  because the r.h.s. is zero for $y>1.$ The reason for  this is in absence of the residues to the right from the straight vertical line in the complex $z$ plane over which the integral  (\ref{MMT Cauchy})  is taken as it has been explained in Subsection \ref{MMG}.  However,  coincidence of integrals does not mean  coincidence of integrands \cite{Kondrashuk:2000br,Cvetic:2002dx} even if the integration domains coincide.  Indeed, we may use a known identity \cite{Faouzi}
\begin{eqnarray} \label{sin}
\G(z)\G(1-z) = \frac{\pi}{\sin{\pi z}}.
\end{eqnarray}
This would mean for the r.h.s. of (\ref{vabene})
\begin{eqnarray*} 
 \G\le\frac{5}{3}\ri  \int\limits_{\delta-i\infty}^{\delta + i\infty} \frac{y^{-z}\G(z)}{\G\le z+ \frac{5}{3} \ri}  ~dz \\
 =  \G\le\frac{5}{3}\ri \int\limits_{\delta-i\infty}^{\delta + i\infty} \frac{y^{-z}\G(z)  \G\le-\frac{2}{3} -z\ri  \sin \pi\le z+ \frac{5}{3}  \ri  }{\pi}  ~dz \\
=  \frac{1}{\G\le-\frac{2}{3} \ri} 
 \int\limits_{\delta-i\infty}^{\delta + i\infty} \frac{y^{-z}\G(z)  \G\le-\frac{2}{3} -z\ri  \sin \pi\le z+ \frac{5}{3}  \ri  }
{  \sin  \frac{5\pi}{3}  }  ~dz \\
=  \frac{1}{\G\le-\frac{2}{3} \ri} 
 \int\limits_{\delta-i\infty}^{\delta + i\infty} y^{-z}\G(z)  \G\le-\frac{2}{3} -z\ri  \le \sin \pi z \cot \frac{5\pi}{3} \right.  \no  \\ 
+ \left.  \cos \pi z  \ri  ~dz.
\end{eqnarray*}
Thus, the factor $\sin \pi z \cot \frac{5\pi}{3} +  \cos \pi z$ kills the contributions of the ``right'' poles
produced by the factor $\G\le-\frac{2}{3} -z\ri$ in the integrand when we close the contour to the right complex infinity for $y>1.$ However for the ``left'' residues, coming form $\G(z),$ the factor
$\cos \pi z$ reproduces the factor $(-1)^z$ on the singular points when we close the contour to the left complex infinity for $y \in [0,1]$  and then we obtain the same residues as they would be obtained from 
the l.h.s. of (\ref{vabene}),
\begin{eqnarray*} 
(1-y)^{\frac{2}{3} }  = \frac{1}{2i\pi}  \frac{1}{\G\le-\frac{2}{3} \ri} \oint\limits_{C}  (-y)^{-z} \G(z)\G\le-\frac{2}{3} -z\ri dz,
 \end{eqnarray*}
with the contour $C$ closed to the left complex infinity.

This simple example helps to find a modification of the inverse integral transformation of the Mellin moments
in more difficult cases considered in the next Subsection \ref{another} when we do not have an infinite radius of convergence for the transforming function $F(y),$  too.

Thus, the difference between the standard inverse transformation of Mellin moments and the inverse MB transformations may be extracted to this identity valid for $0<y<1,$
\begin{eqnarray} \label{vabene-2}
(1-y)^\frac{2}{3}  = \frac{1}{2i\pi}  \frac{1}{\G\le-\frac{2}{3} \ri} \oint\limits_{C}  (-y)^{-z} \G(z)\G\le-\frac{2}{3} -z\ri dz \no\\
= \frac{1}{2i\pi} \frac{1}{\G\le-\frac{2}{3} \ri} \oint\limits_{C} y^{-z}\G(z)  \G\le-\frac{2}{3} -z\ri
\le \sin \pi z \cot \frac{5\pi}{3} \right. \no\\
+  \left.  \cos \pi z  \ri  ~dz.
\end{eqnarray}
The contour $C$ contains the straight vertical line of (\ref{vabene}) and the left complex infinity. The integrands on the l.h.s. and on r.h.s. do not coincide however the ``left'' residues completely coincide. By Cauchy integral formula the results for integrals coincide in this case.

The purpose of this paper is to generalize the inverse integral transformation of the Mellin moments in such a way that the original function $F(y)$ is recovered for any value of its argument $y$ even if $y>1$. As we have mentioned already at the beginning of this Section  \ref{examples}, it is not a problem when $F(y)$
has an infinite radius of convergence at the point $y=0$. However, in the present case
$F(y) = (1-y)^\frac{2}{3}$ the radius of convergence is 1. The inverse MB transformation (\ref{moment-barnes-2})
recovers the same result even if $y>1.$ In such a case the contour $C$ should be closed to  the right complex infinity,
\begin{eqnarray} \label{barnes-3}
F(y)  =  (1-y)^\frac{2}{3} = y^\frac{2}{3}\le 1 - \frac{1}{y} \ri^\frac{2}{3}  \no\\
 =  y^\frac{2}{3}\sum_{k=0}^{\infty}\frac{\Gamma\le-\frac{2}{3}+k\ri}{\Gamma\le-\frac{2}{3}\ri}\frac{y^{-k}}{k!}\no\\
\frac{1}{2i\pi} \frac{1}{\G\le-\frac{2}{3} \ri} \int\limits_{\delta-i\infty}^{\delta + i\infty}  (-y)^{z}\G(z)  \G\le-\frac{2}{3} -z\ri  ~dz,
\end{eqnarray}
so that we have the convergent series in terms of negative powers of $y.$ The inverse MB transformation seems to be enough for the purpose to recover the original function for any value of $y$ but the lack of such a representation is that the contour remains open line initially but we would prefer from the very beginning to have a closed contour for the inverse transformation of the Mellin moments like we have in Theorem \ref{Moment-2} for the case of an infinite radius of convergence. For this purpose we consider the integral on the r.h.s. of (\ref{vabene-2}) for $y>1,$
\begin{eqnarray} \label{ha}
\frac{1}{2i\pi} \frac{1}{\G\le-\frac{2}{3} \ri} \oint\limits_{C} y^{-z}\G(z)  \G\le-\frac{2}{3} -z\ri
\le \sin \pi z \cot \frac{5\pi}{3} \right. \no\\
+ \left.   \cos \pi z  \ri  ~dz
=  \sum_{k=0}^{\infty}\frac{\Gamma\le-\frac{2}{3}+k\ri}{\Gamma\le-\frac{2}{3}\ri}\frac{y^k}{k!}.
\end{eqnarray}
The result is different comparing with the inverse MB transformation (\ref{barnes-3}) because the series obtained from the calculation of the residues inside this closed contour is divergent for $y>1$ because the contour $C$
is closed to the left complex infinity for $y>1.$ Apparently, the only contribution from the contour $C$
corresponds to the left complex infinity which is equal to this divergent series according to the construction done
in Subsection \ref{MMG},
\begin{eqnarray*} 
\frac{1}{2i\pi} \frac{1}{\G\le-\frac{2}{3} \ri} \oint\limits_{C} y^{-z}\G(z)  \G\le-\frac{2}{3} -z\ri
\le \sin \pi z \cot \frac{5\pi}{3}  \right. \\
+  \left. \cos \pi z  \ri  ~dz
=  \sum_{k=0}^{\infty}\frac{\Gamma\le-\frac{2}{3}+k\ri}{\Gamma\le-\frac{2}{3}\ri}\frac{y^k}{k!}  \\
= \frac{\G\le\frac{5}{3}\ri}{2i\pi}  \oint\limits_C \frac{y^{-z}\G(z)}{\G\le z+ \frac{5}{3} \ri}  ~dz
= \frac{\G\le\frac{5}{3}\ri}{2i\pi}  \int\limits_{-\infty+i\infty}^{-\infty - i\infty} \frac{y^{-z}\G(z)}{\G\le z+ \frac{5}{3} \ri}  ~dz.
\end{eqnarray*}
By using the MB inverse transformation (\ref{moment-barnes-2}) for $y>1$ we may write for the same closed contour $C$ that
\begin{eqnarray*}
 (1-y)^\frac{2}{3} = \frac{1}{2i\pi} \frac{1}{\G\le-\frac{2}{3} \ri} \int\limits_{\delta-i\infty}^{\delta + i\infty} (-y)^{-z}\G(z)  \G\le-\frac{2}{3} -z\ri    ~dz  \\
= \frac{1}{2i\pi} \frac{1}{\G\le-\frac{2}{3} \ri} \oint\limits_{C} (-y)^{-z}\G(z)  \G\le-\frac{2}{3} -z\ri~dz\\
- \frac{1}{2i\pi} \frac{1}{\G\le-\frac{2}{3} \ri} \int\limits_{-\infty+i\infty}^{-\infty - i\infty} (-y)^{-z}\G(z)  \G\le-\frac{2}{3} -z\ri    ~dz \\
=  \sum_{k=0}^{\infty}\frac{\Gamma\le-\frac{2}{3}+k\ri}{\Gamma\le-\frac{2}{3}\ri}\frac{y^k}{k!} \\
- \frac{1}{2i\pi} \frac{1}{\G\le-\frac{2}{3} \ri} \int\limits_{-\infty+i\infty}^{-\infty - i\infty} (-y)^{-z}\G(z)  \G\le-\frac{2}{3} -z\ri    ~dz.
\end{eqnarray*}

Thus, we may write for the divergent series (when $y>1$) this chain of identities
\begin{eqnarray*} 
 \sum_{k=0}^{\infty}\frac{\Gamma\le-\frac{2}{3}+k\ri}{\Gamma\le-\frac{2}{3}\ri}\frac{y^k}{k!}  
= \frac{\G\le\frac{5}{3}\ri}{2i\pi}  \int\limits_{-\infty+i\infty}^{-\infty - i\infty} \frac{y^{-z}\G(z)}{\G\le z+ \frac{5}{3} \ri}  ~dz \\
= \frac{1}{2i\pi} \frac{1}{\G\le-\frac{2}{3} \ri} \int\limits_{-\infty+i\infty}^{-\infty - i\infty}     y^{-z}\G(z)  \G\le-\frac{2}{3} -z\ri
\le \sin \pi z  \right. \\
\left. \cot \frac{5\pi}{3}  +    \cos \pi z  \ri  ~dz \\
=  (1-y)^\frac{2}{3} \\
+ \frac{1}{2i\pi} \frac{1}{\G\le-\frac{2}{3} \ri} \int\limits_{-\infty+i\infty}^{-\infty - i\infty} (-y)^{-z}\G(z)  \G\le-\frac{2}{3} -z\ri    ~dz.
 \end{eqnarray*}
Thus, even if we work with the divergent series and integrals  at the intermediate steps, we come to a correct relation
for the finite integrals
\begin{eqnarray} \label{Y>1}
(1-y)^\frac{2}{3} \no \\
= \frac{1}{2i\pi} \frac{1}{\G\le-\frac{2}{3} \ri} \int\limits_{-\infty+i\infty}^{-\infty - i\infty} y^{-z}
\G(z) \G\le-\frac{2}{3} -z\ri \no\\
\le \sin \pi z \cot \frac{5\pi}{3} +  \cos \pi z  - (-1)^{-z} \ri   ~dz,
 \end{eqnarray}
which  is valid for any $y \in [1,\infty[.$ At the same time, for $y \in [0,1]$ this integral (\ref{Y>1})
is zero by construction.

We may re-write (\ref{Y>1}) as a contour integral
\begin{eqnarray} \label{y>>1}
(1-y)^\frac{2}{3} \no \\
= \frac{1}{2i\pi} \frac{1}{\G\le-\frac{2}{3} \ri} \oint\limits_{C_2} y^{-z}\G(z) \G\le-\frac{2}{3} -z\ri \no\\
\le \sin \pi z \cot \frac{5\pi}{3} +  \cos \pi z  - (-1)^{-z} \ri   ~dz
 \end{eqnarray}
such that all the poles of both Euler gamma functions $\G(z)$ and  $\G\le-\frac{2}{3} -z\ri$ are inside
the contour $C_2.$ This is the only condition for the contour $C_2.$  We cannot apply the closed contour integral
representation (\ref{y>>1}) to the case $y \in [0,1]$ because the open contour (\ref{Y>1}) cannot be closed to the right complex infinity which would contribute in such a case. The open contour  (\ref{Y>1}) should be closed on the straight vertical line (\ref{vabene}) in order to construct a closed rectangular contour, after this result for (\ref{Y>1})
remains zero in the case $y \in [0,1].$

At the same time, it is clear that
in  (\ref{vabene-2}) the contour $C$ may be replaced with this contour $C_2,$ too. Thus, for any $y \in [0,1]$
we may write
\begin{eqnarray} \label{Y<1}
(1-y)^\frac{2}{3} \no\\
= \frac{1}{2i\pi} \frac{1}{\G\le-\frac{2}{3} \ri} \oint\limits_{C_2} y^{-z}\G(z)  \G\le-\frac{2}{3} -z\ri
\le \sin \pi z \cot \frac{5\pi}{3} \right. \no\\
+  \left.  \cos \pi z  \ri  ~dz.
\end{eqnarray}
Combining (\ref{y>>1}) and  (\ref{Y<1}) together, we come to the representation which is valid for all the values  $y \in [0,\infty[,$
\begin{eqnarray} \label{y}
(1-y)^\frac{2}{3}
= \frac{1}{2i\pi} \frac{1}{\G\le-\frac{2}{3} \ri} \int\limits_{-\infty+i\infty}^{-\infty - i\infty} y^{-z}
\G(z) \no\\
\G\le-\frac{2}{3} -z\ri
\le \sin \pi z \cot \frac{5\pi}{3} +  \cos \pi z  \ri   ~dz \no\\
- \frac{1}{2i\pi} \frac{1}{\G\le-\frac{2}{3} \ri} \int\limits_{-\infty+i\infty}^{-\infty - i\infty} (-y)^{-z}
\G(z) \G\le-\frac{2}{3} -z\ri dz \no\\
+ \frac{1}{2i\pi} \frac{1}{\G\le-\frac{2}{3} \ri} \int\limits_{\delta-i\infty}^{\delta + i\infty} y^{-z}
\G(z) \no \\
\G\le-\frac{2}{3} -z\ri
\le \sin \pi z \cot \frac{5\pi}{3} +  \cos \pi z  \ri   ~dz \no\\
=  \frac{1}{2i\pi} \frac{1}{\G\le-\frac{2}{3} \ri} \oint\limits_{C_2} y^{-z}
\G(z) \no\\
\G\le-\frac{2}{3} -z\ri
\le \sin \pi z \cot \frac{5\pi}{3} +  \cos \pi z  \ri   ~dz \no\\
- \frac{1}{2i\pi} \frac{1}{\G\le-\frac{2}{3} \ri} \int\limits_{-\infty+i\infty}^{-\infty - i\infty} (-y)^{-z}
\G(z) \G\le-\frac{2}{3} -z\ri dz.
\end{eqnarray}

The advantage of this representation is that it is an integral over closed contour from the very beginning but not over an open line like in the MB transformation (\ref{moment-barnes-2}).

\subsection{Another example of $z$-moment with infinitely many poles} \label{another}

We have checked in the previous Subsections \ref{poly}  and \ref{infinite} that  Theorem \ref{Moment-2} is valid for  $z$-moments  with
a finite number of poles  in the complex $z$-plane of its  variable or for $z$-moments possessing infinitely many poles in this plane  if the  convergence radius 
is bigger than 1.  In case if the radius of convergence is 1 or less than 1 we need to find a way to work with  divergent series and to modify the integrands
of the inverse integral transformation of the Mellin moments so that these inverse transformations  would recover the original function  $F(y)$
 for both the domains  $y \in [0,1]$ and  $y \in [1,\infty[$  maintaining the integration contour  closed from the very beginning and without modifying this contour until evaluation of  the contour integral is finished.  The example of such a modification of  integrand is given
in Subsection \ref{comparing}.   In this Subsection we consider a similar example  $F(y)  =  (1-y)^\ve$  in which $\ve$ is small real positive. 
Its Mellin moment is again the  Euler beta function,   
\begin{eqnarray} \label{beta-3} 
\int\limits_{0}^{1} y^{z-1}(1-y)^\ve dy = \frac{\G(z)\G\le 1 +\ve\ri}{\G\le z+ 1 +\ve \ri}. 
\end{eqnarray}

We follow the lines of the previous Subsection \ref{comparing} applying the standard  inverse integral transformation  (\ref{MMT Cauchy}) of the Mellin moment and returning  the original function $F(y)  =  (1-y)^\ve$ after such a transformation
\begin{eqnarray}\label {moment-barnes-50}
 \frac{\G\le1+\ve\ri}{2i\pi}  \int\limits_{\delta-i\infty}^{\delta + i\infty} \frac{y^{-z}\G(z)}{\G\le z+ 1 +\ve \ri}  ~dz  \\
 =  \sum_{k=0}^{\infty}\frac{\Gamma\le-\ve+k\ri}{\Gamma\le-\ve\ri}\frac{y^k}{k!}  = (1-y)^\ve \no.
\end{eqnarray}

According to the construction done in Subsection \ref{MMG} the contour should be closed to the left complex infinity because we suppose  in (\ref{moment-barnes-50})
the standard domain $0<y<1$ for the Mellin moments. Then we calculate the residues inside the contour.   The first two residues are
\begin{eqnarray*}
\res_{z= 0} \frac{y^{-z}\G(z)}{\G\le z+ 1 +\ve \ri}  =    \frac{1}{\G\le 1 + \ve \ri}, \\
\res_{z= -1} \frac{y^{-z}\G(z)}{\G\le z+ 1 +\ve   \ri}  =    -\frac{y}{\G\le  \ve  \ri}.
\end{eqnarray*}

The binomial expansion (\ref{binom}) is used on the right hand side of (\ref{moment-barnes-50})
for $y \in [0,1].$    Eq. (\ref{moment-barnes-50})  in turn may be proved from this MB representation
\begin{eqnarray} \label {moment-barnes-51}
(1-y)^\ve \no\\
= \frac{1}{2i\pi}  \frac{1}{\G\le-\ve \ri} \int\limits_{\delta-i\infty}^{\delta + i\infty}(-y)^{-z} \G(z)\G\le-\ve -z\ri dz 
\end{eqnarray}
 which  is valid for any real $y \in [0,\infty[.$  By comparing (\ref{moment-barnes-50})  and  (\ref{moment-barnes-51})  
we conclude that for $y \in [0,1]$
\begin{eqnarray}  \label{killing}  
\frac{1}{2i\pi}  \frac{1}{\G\le-\ve \ri} \int\limits_{\delta-i\infty}^{\delta + i\infty}(-y)^{-z} \G(z)\G\le-\ve -z\ri dz  \no\\ 
= (1-y)^\ve  =  \frac{\G\le1+\ve\ri}{2i\pi}  \int\limits_{\delta-i\infty}^{\delta + i\infty} 
\frac{y^{-z}\G(z)}{\G\le z+ 1 +\ve \ri}  ~dz.
\end{eqnarray}
This identity is not valid for  $y \in [1,\infty[$  because the r.h.s. is zero for $y>1.$ The reason for  this is in absence of the residues to the right from the straight vertical line in the complex $z$ plane over which the integral  (\ref{MMT Cauchy})  is taken as it has been explained in Subsection \ref{MMG}.  However, the integrands are not identical. By using the identity (\ref{sin}) we come to 
\begin{eqnarray*} 
\G\le1+\ve\ri  \int\limits_{\delta-i\infty}^{\delta + i\infty}   \frac{y^{-z}\G(z)}{\G\le z+ 1 +\ve \ri}  ~dz \no \\
=  \G\le1+\ve\ri   \int\limits_{\delta-i\infty}^{\delta + i\infty} \frac{y^{-z}\G(z)\G\le -z-\ve \ri \sin \pi\le z+ 1+\ve  \ri}{\pi}  
~dz  \\
=    \frac{1}{\G\le-\ve \ri} \int\limits_{\delta-i\infty}^{\delta + i\infty} y^{-z}\G(z)\G\le -z-\ve \ri 
\le \sin \pi z \cot \pi(1+\ve)  \right. \\ \left.
+  \cos \pi z  \ri~dz . 
\end{eqnarray*}
Thus, the factor $ \sin \pi z \cot \pi(1+\ve)  +  \cos \pi z  $ kills the contributions of the ``right'' poles
produced by the factor $\G\le-\ve -z\ri$ in the integrand when we close the contour to the right complex infinity for $y>1.$ However for the ``left'' residues, coming form $\G(z),$ the factor
$\cos \pi z$ reproduces the factor $(-1)^z$ on the singular points when we close the contour to the left complex infinity for $y \in [0,1]$  and then we obtain the same residues as they would be obtained from the l.h.s. of (\ref{killing}),
\begin{eqnarray*} 
(1-y)^{\ve } = \frac{1}{2i\pi} \frac{1}{\G\le-\ve \ri} \oint\limits_C(-y)^{-z} \G(z)\G\le-\ve -z\ri dz,
\end{eqnarray*}
with the contour $C$ closed to the left complex infinity.

The difference between the standard inverse transformation of Mellin moments and the inverse MB transformations may be extracted at the level of closed contours  to this identity valid for $0<y<1,$
\begin{eqnarray}  \label{killing-2} 
(1-y)^{\ve }  =   \frac{1}{2i\pi}  \frac{1}{\G\le-\ve \ri} \oint\limits_{C}(-y)^{-z} \G(z)\G\le-\ve -z\ri dz \no\\
=    \frac{1}{2i\pi}  \frac{1}{\G\le-\ve \ri} \oint\limits_{C} y^{-z}\G(z)\G\le -z-\ve \ri  \no\\
\le \sin \pi z \cot \pi(1+\ve)  +  \cos \pi z  \ri~dz.  
\end{eqnarray}
The contour $C$ contains the straight vertical line of (\ref{killing}) and the left complex infinity. The integrands on the l.h.s. and on r.h.s. do not coincide however the ``left'' residues (produced by $\G(z)$) completely coincide. By Cauchy integral formula the results for integrals coincide in this case.

The purpose of this paper is to generalize the inverse integral transformation of the Mellin moments in such a way that the original function $F(y)$ is recovered for any value of its argument $y$ even if $y>1$. As we have mentioned already at the beginning of this Section  \ref{examples}, this may be done immediately  when $F(y)$
has an infinite radius of convergence at the point $y=0$. However, in the present case
$F(y) = (1-y)^\ve$ the radius of convergence is 1. The inverse MB transformation (\ref{moment-barnes-51})
recovers the same result even if $y>1.$ In such a case the contour $C$ should be closed to  the right complex infinity,
\begin{eqnarray}  \label{barnes-4}  
F(y)  =  (1-y)^\ve = (-y)^\ve \le 1 - \frac{1}{y} \ri^\ve \no \\
= \frac{1}{2i\pi} \frac{(-y)^{\ve}}{\G\le-\ve \ri} \int\limits_{\delta-i\infty}^{\delta + i\infty}  (-y)^{z}\G(z)  \G\le-\ve -z\ri  ~dz  \no\\
=  (-y)^\ve\sum_{k=0}^{\infty}\frac{\Gamma\le-\ve+k\ri}{\Gamma\le-\ve\ri}\frac{y^{-k}}{k!},
\end{eqnarray}
so that we have the convergent series in terms of the negative powers of $y.$ The inverse MB transformation seems to be enough for the purpose to recover the original function for any value of $y$ but the lack of such a representation is that the contour remains an open line initially but we would prefer from the very beginning to have a closed contour for the inverse transformation of the Mellin moments like we have in Theorem \ref{Moment-2} for the case of an infinite radius of convergence. For this purpose we consider the integral on the r.h.s. of (\ref{killing-2}) for $y>1,$
\begin{eqnarray} \label{nat} 
   \frac{1}{2i\pi} \frac{1}{\G\le-\ve \ri} \oint\limits_C y^{-z}\G(z)\G\le -z-\ve \ri
\le \sin \pi z \cot \pi(1+\ve) \right. \no\\
\left. +  \cos \pi z  \ri~dz
=  \sum_{k=0}^{\infty}\frac{\Gamma\le-\ve+k\ri}{\Gamma\le-\ve\ri}\frac{y^k}{k!} .
\end{eqnarray}
The result is different comparing with the inverse MB transformation (\ref{barnes-4}) because the series obtained from the calculation of the residues inside this closed contour is divergent for $y>1$ because the contour $C$
is closed to the left complex infinity for $y>1.$ Apparently, the only contribution from the contour $C$
corresponds to the left complex infinity which is equal to this divergent series according to the construction done
in Subsection \ref{MMG},
\begin{eqnarray*} 
 \frac{1}{\G\le-\ve \ri} \frac{1}{2i\pi} \oint\limits_C y^{-z}\G(z)\G\le -z-\ve \ri 
\le \sin \pi z \cot \pi(1+\ve)  \right. \\ 
+ \left.   \cos \pi z  \ri~dz  =  \sum_{k=0}^{\infty}\frac{\Gamma\le-\ve+k\ri}{\Gamma\le-\ve\ri}\frac{y^k}{k!}   \\
=  \frac{\G\le1+\ve\ri}{2i\pi}  \oint\limits_{C} \frac{y^{-z}\G(z)}{\G\le z+ 1 +\ve \ri}  ~dz \\ 
=  \frac{\G\le1+\ve\ri}{2i\pi}  \int\limits_{-\infty+i\infty}^{-\infty - i\infty} \frac{y^{-z}\G(z)}{\G\le z+ 1 +\ve \ri}  ~dz. 
\end{eqnarray*}
By using the MB inverse transformation (\ref{moment-barnes-51}) for $y>1$ we may write for the same closed contour $C$ that
\begin{eqnarray*} 
(1-y)^\ve   
=    \frac{1}{\G\le-\ve \ri} \frac{1}{2i\pi}\int\limits_{\delta-i\infty}^{\delta + i\infty} (-y)^{-z}\G(z)\G\le -z-\ve \ri ~dz \\ 
 = \frac{1}{\G\le-\ve \ri} \frac{1}{2i\pi}\oint\limits_C(-y)^{-z}\G(z)\G\le -z-\ve \ri ~dz \\
-  \frac{1}{\G\le-\ve \ri}\frac{1}{2i\pi} \int\limits_{-\infty+i\infty}^{-\infty - i\infty} (-y)^{-z}\G(z)\G\le -z-\ve \ri dz \\
=  \sum_{k=0}^{\infty}\frac{\Gamma\le-\ve+k\ri}{\Gamma\le-\ve\ri}\frac{y^k}{k!}  \\
-  \frac{1}{\G\le-\ve \ri}\frac{1}{2i\pi} \int\limits_{-\infty+i\infty}^{-\infty - i\infty} (-y)^{-z}\G(z)\G\le -z-\ve \ri dz.
\end{eqnarray*}

Thus, we may write for the divergent series (when $y>1$) this chain of identities
\begin{eqnarray*} 
\sum_{k=0}^{\infty}\frac{\Gamma\le-\ve+k\ri}{\Gamma\le-\ve\ri}\frac{y^k}{k!}  
=  \frac{\G\le1+\ve\ri}{2i\pi}  \int\limits_{-\infty+i\infty}^{-\infty - i\infty} \frac{y^{-z}\G(z)}{\G\le z+ 1 +\ve \ri}  ~dz  \\
  =    \frac{1}{\G\le-\ve \ri} \frac{1}{2i\pi} \int\limits_{-\infty + i\infty}^{-\infty - i\infty}y^{-z}\G(z) \no \\
  \G\le -z-\ve \ri \le \sin \pi z \cot \pi(1+\ve)  +  \cos \pi z  \ri~dz  \\
=  (1-y)^\ve     +  \frac{1}{\G\le-\ve \ri}\frac{1}{2i\pi} \int\limits_{-\infty+i\infty}^{-\infty - i\infty} (-y)^{-z}\G(z)\G\le -z-\ve \ri dz.
 \end{eqnarray*}

Thus, even if we work with the divergent series and integrals  at the intermediate steps, we come to a correct relation
for the finite integrals
\begin{eqnarray} \label{y>12} 
(1-y)^\ve  = \frac{1}{\G\le-\ve \ri} \frac{1}{2i\pi} \int\limits_{-\infty + i\infty}^{-\infty - i\infty}y^{-z}\G(z)\G\le -z-\ve \ri \no\\
 \le \sin \pi z \cot \pi(1+\ve)  +  \cos \pi z  - (-1)^{-z} \ri~dz,
 \end{eqnarray}
which  is valid for any $y \in [1,\infty[.$ At the same time, for $y \in [0,1]$ this integral (\ref{y>12})
is zero by construction.

The construction (\ref{y>12}) can be  written  as a contour integral
\begin{eqnarray} \label{y>>12} 
(1-y)^\ve  = \frac{\G\le1+\ve\ri}{2i\pi}  \int\limits_{-\infty+i\infty}^{-\infty - i\infty} \frac{y^{-z}\G(z)}{\G\le z+ 1 +\ve \ri}  ~dz \no \\
 -   \frac{1}{\G\le-\ve \ri}\frac{1}{2i\pi} \int\limits_{-\infty+i\infty}^{-\infty - i\infty} (-y)^{-z}\G(z)\G\le -z-\ve \ri dz \no\\
= \frac{\G\le1+\ve\ri}{2i\pi}  \int\limits_{-\infty+i\infty}^{-\infty - i\infty} \frac{y^{-z}\G(z)}{\G\le z+ 1 +\ve \ri}  ~dz  \no\\
 -   \frac{\G\le1+\ve \ri }{2i\pi} \int\limits_{-\infty+i\infty}^{-\infty - i\infty} \frac{ (-y)^{-z} \G(z)}{\G\le z+ 1 +\ve \ri}   
\frac{\sin{\pi \ve}}{\sin{\pi(z+\ve)}}dz \no\\
= \frac{\G\le1+\ve\ri}{2i\pi}  \int\limits_{-\infty+i\infty}^{-\infty - i\infty} \frac{y^{-z}\G(z)}{\G\le z+ 1 +\ve \ri} 
\left[1  \right.\no\\
\left.  -  \frac{(-1)^{-z}\sin{\pi \ve}}{\sin{\pi(z+\ve)}}  \right]   ~dz \no\\
= \frac{\G\le1+\ve\ri}{2i\pi}  \oint\limits_{C_3} \frac{y^{-z}\G(z)}{\G\le z+ 1 +\ve \ri} 
\left[1 -  \frac{(-1)^{-z}\sin{\pi \ve}}{\sin{\pi(z+\ve)}}  \right]   ~dz,
\end{eqnarray}
such that all the poles of both Euler gamma functions $\G(z)$ and  $\G\le-\ve -z\ri$ are inside
the contour $C_3.$ This is the only condition for the contour $C_3.$ In contrast to (\ref{y>12})
we cannot apply the contour integral (\ref{y>>12}) to  $y \in [0,1]$ because in this case a divergent
series would arise from the residues on the right complex half-plane. The integral over the contour $C_3$ is valid
only for  $y \in [1,\infty[.$

At the same time, it is clear that
in in (\ref{killing-2}) the contour $C$ may be taken as this contour $C_3,$ too. Thus, for any $y \in [0,1]$
we may write
\begin{eqnarray} \label{y<12}
(1-y)^\ve  = \frac{\G\le1+\ve\ri}{2i\pi}  \oint\limits_{C_3} \frac{y^{-z}\G(z)}{\G\le z+ 1 +\ve \ri} ~dz .
\end{eqnarray}
Combining (\ref{y>>12}) and  (\ref{y<12}) together, we come to the representation which is valid for all the values  $y \in [0,\infty[,$
\begin{eqnarray} \label{yy}
(1-y)^\ve  = \frac{\G\le1+\ve\ri}{2i\pi}  \int\limits_{-\infty+i\infty}^{-\infty - i\infty} \frac{y^{-z}\G(z)}{\G\le z+ 1 +\ve \ri}  ~dz \no\\
-   \frac{1}{\G\le-\ve \ri}\frac{1}{2i\pi} \int\limits_{-\infty+i\infty}^{-\infty - i\infty} (-y)^{-z}\G(z)\G\le -z-\ve \ri dz \no\\
+ \frac{\G\le1+\ve\ri}{2i\pi}  \int\limits_{\delta - i\infty}^{\delta + i\infty}
\frac{y^{-z}\G(z)}{\G\le z+ 1 +\ve \ri}  ~dz \no \\
=  \frac{\G\le1+\ve\ri}{2i\pi}  \oint\limits_{C_3} \frac{y^{-z}\G(z)}{\G\le z+ 1 +\ve \ri}  ~dz \no\\
-   \frac{1}{\G\le-\ve \ri}\frac{1}{2i\pi} \int\limits_{-\infty+i\infty}^{-\infty - i\infty} (-y)^{-z}\G(z)\G\le -z-\ve \ri dz.
\end{eqnarray}
The advantage of this representation is that  $C_3$ is an integral over a closed contour from the very beginning but it is not an open contour like in the MB transformation (\ref{moment-barnes-51}).

\subsection{Euler digamma function as $z$-moment} \label{Sub-psi}

In the present Subsection \ref{Sub-psi} we consider another example of the transforming function for which the radius of convergence is limited, $F(y) = \ln(1-y).$ The Mellin moment of this function appears in applications to quantum chromodynamics \cite{Alvarez:2016juq}, in particular in calculation of splitting functions \cite{Alvarez:2016juq}. These functions are commented in the next Section \ref{Sec-pro}. This is why we
consider it in details here. The results given here are straightforward consequences of the results obtained
in the previous Subsection \ref{another}. They may be obtained from the previous example $(1-y)^\ve$
by taking derivative with respect to $\ve,$
\begin{eqnarray*} 
\left.\left(\frac{d}{d\ve}\int\limits_{0}^{1} y^{z-1}(1-y)^\ve dy\right)\right|_{\ve=0} = \left.\left(\frac{d}{d\ve}\frac{\G(z)\G\le 1 +\ve\ri}{\G\le z+ 1 +\ve \ri}\right)\right|_{\ve=0}  \\
=  \left.\left(\frac{\G(z)\G\le 1 +\ve\ri\le\psi(1+\ve) - \psi(1+\ve+z) \ri}{\G\le z+ 1 +\ve \ri}\right)\right|_{\ve=0} \\
= \frac{\psi(1) - \psi(1+z) }{ z}.
\end{eqnarray*}
We come to this expression for the Mellin moment of this function, which may be expressed via Euler digamma (psi)  function,
\begin{eqnarray*} 
\int\limits_{0}^{1} y^{z-1}\ln{(1-y)}~ dy = -\sum_{k=1}^{\infty}  \int\limits_{0}^{1} \frac{y^{z+k-1}}{k}~ dy \\
= - \sum_{k=1}^{\infty}  \frac{1}{k(z+k)}  = - \frac{1}{z}\sum_{k=1}^{\infty}  \le  \frac{1}{k} - \frac{1}{z+k} \ri \\
= \frac{\psi(1) - \psi(1+z) }{ z}. 
\end{eqnarray*}
This formula is one of the known representations for the $\psi$ function \cite{Alvarez:2016juq},
\begin{eqnarray} \label{psi} 
\psi(z) = \sum_{k=1}^{\infty}\le \frac{1}{k} - \frac{1}{k+z-1} \ri  + \psi(1)
\end{eqnarray}
This representation shows infinite number of poles corresponding to the non-positive integers in the complex
$z$ plane of the variable of the Mellin moment.

The standard inverse integral transformation (\ref{MMT Cauchy}) for the Mellin moment recovers the original function
\begin{eqnarray}\label {moment-barnes-50-2}
 \frac{1}{2i\pi}  \int\limits_{\delta-i\infty}^{\delta + i\infty} \frac{y^{-z}\le \psi(1) - \psi(1+z)  \ri}{z}   =  \ln{(1-y)}. 
\end{eqnarray}
We reproduce it immediately by  residue calculus for this function after closing the contour to the left complex infinity according to the theory described in Section \ref{MMB} for the standard domain   $0<y<1.$
The singularities of the $\psi(z)$ are explicit in the representation (\ref{psi}). This representation (\ref{psi}) does not show
poles to the right from the straight vertical line over which the integral is taken as it should be according to
the standard theory of the Mellin moments so that the integral (\ref{moment-barnes-50-2}) becomes zero for the
domain $y>1.$ This standard theory in described in Section  \ref{MMB}.

The MB transformation (\ref{MBG}) needs a bit different evaluation in this case. We take a derivative of the formula (\ref {moment-barnes-51}) with respect to $\ve,$
\begin{eqnarray} \label {moment-barnes-51-2}
\ln{(1-y)} = \frac{1}{2i\pi}  \lim_{\ve \rar 0}\frac{1}{\G\le-\ve\ri}\int\limits_{\delta-i\infty}^{\delta + i\infty}(-y)^{-z} \G(z)\no\\
\G\le -\ve-z\ri\le \psi(-\ve) -\psi(-\ve-z)\ri dz \no\\
= \frac{1}{2i\pi}  \lim_{\ve \rar 0}\G\le1+\ve\ri\sin{\pi\ve} \int\limits_{\delta-i\infty}^{\delta + i\infty}(-y)^{-z} \G(z)\no\\
\frac{ \psi(-\ve) -\psi(-\ve-z)}{\G\le1+\ve+z  \ri \sin{\pi(\ve+z)}}  dz \no\\
= \frac{1}{2i\pi}  \lim_{\ve \rar 0}\sin{\pi\ve} \int\limits_{\delta-i\infty}^{\delta + i\infty}
\frac{(-y)^{-z}  \le\psi(-\ve) -\psi(-\ve-z)\ri}{z\sin{\pi(\ve+z)}}  dz  \no\\
= \frac{1}{2i}   \int\limits_{-1+\delta-i\infty}^{-1+\delta + i\infty}\frac{(-y)^{-z}}{z\sin{\pi z }}  dz \no\\
=   \frac{1}{2i\pi}   \int\limits_{-1+\delta-i\infty}^{-1+\delta + i\infty}\frac{(-y)^{-z}\G(z)\G(1-z)}{z}  dz.
\end{eqnarray}
We use the formula (\ref{sin}) to take the limit. The residue at the point $z=0$ disappears. This allows us to shift the contour and take a limit for the rest of residues \cite{Diaz:2024tdp}. As we have commented in the previous Subsections \ref{another} and \ref{comparing}, the MB transformation is valid for any real $y \in [0,\infty[$  while the standard inverse transformation of the Mellin moment  recovers zero
for the domain $y \in [1,\infty[.$ By comparing (\ref{moment-barnes-50-2})  and  (\ref{moment-barnes-51-2})
we conclude that for $y \in [0,1]$
\begin{eqnarray}  \label{killing-22}  
\frac{1}{2i\pi}   \int\limits_{-1+\delta-i\infty}^{-1+\delta + i\infty}\frac{(-y)^{-z}\G(z)\G(1-z)}{z}  dz 
=  \ln{(1-y)} \no\\  
= \frac{1}{2i\pi}  \int\limits_{\delta-i\infty}^{\delta + i\infty} \frac{y^{-z}\le \psi(1) - \psi(1+z)  \ri}{z}dz \no\\
= \frac{1}{2i\pi}  \int\limits_{-1+\delta-i\infty}^{-1+\delta + i\infty} \frac{y^{-z}\le \psi(1) - \psi(1+z)  \ri}{z}dz.
\end{eqnarray}
This identity is not valid for  $y \in [1,\infty[$  because the r.h.s. is zero for $y>1.$ The reason for  this is in absence of the residues to the right from the straight vertical line in the complex $z$ plane over which the integral  (\ref{MMT Cauchy})  is taken as it has been explained in Subsection \ref{MMG}.  However, the integrands in (\ref{killing-22}) are not
identical.

The difference between the standard inverse transformation of Mellin moments and the inverse MB transformations may be extracted at the level of closed contours  to this identity valid for $0<y<1,$
\begin{eqnarray}  \label{killing-22-2}
\frac{1}{2i\pi}   \oint\limits_{C}\frac{(-y)^{-z}\G(z)\G(1-z)}{z}  dz 
=  \ln{(1-y)} \no\\  
= \frac{1}{2i\pi}  \oint\limits_{C} \frac{y^{-z}\le \psi(1) - \psi(1+z)  \ri}{z}dz.
\end{eqnarray}
The contour $C$ contains the straight vertical line of (\ref{killing-22}) and the left complex infinity. The integrands on the l.h.s. and on r.h.s. do not coincide however the ``left'' residues (produced by $\G(z)$) completely coincide. By Cauchy integral formula the results for integrals coincide in this case.

The purpose of this paper is to generalize the inverse integral transformation of the Mellin moments in such a way that the original function $F(y)$ is recovered for any value of its argument $y$ even if $y>1$. As we have mentioned already at the beginning of this Section  \ref{examples}, this may be done immediately  when $F(y)$
has an infinite radius of convergence at the point $y=0$. However, in the present case
$F(y) = \ln(1-y)$ the radius of convergence is 1. The inverse MB transformation (\ref{moment-barnes-51-2})
recovers the same result even if $y>1.$ In such a case the contour $C$ should be closed to  the right complex infinity,
\begin{eqnarray}  \label{barnes-4-2}  
F(y)  =  \ln(1-y) = \ln(-y) +\ln \le 1 - \frac{1}{y} \ri  \no \\
= \frac{1}{2i\pi}   \int\limits_{-1+\delta-i\infty}^{-1+\delta + i\infty}\frac{(-y)^{-z}\G(z)\G(1-z)}{z}  dz   \no\\
= \ln(-y) -   \sum_{k=1}^{\infty}\frac{y^{-k}}{k}, 
\end{eqnarray}
so that we have the convergent series in terms of the negative powers of $y.$ The inverse MB transformation seems to be enough for the purpose to recover the original function for any value of $y.$ However, the lack of such a representation is that the contour remains an open line initially but we would prefer from the very beginning to have a closed contour for the inverse transformation of the Mellin moments like we have in Theorem \ref{Moment-2} for the case of an infinite radius of convergence. For this purpose we consider the integral on the r.h.s. of (\ref{killing-22}) for $y>1,$
\begin{eqnarray} \label{nat-2} 
  \frac{1}{2i\pi}  \oint\limits_{C} \frac{y^{-z}\le \psi(1) - \psi(1+z)  \ri}{z}dz  =  
- \sum_{k=1}^{\infty}\frac{y^k}{k} .
\end{eqnarray}

The result is different comparing with the inverse MB transformation (\ref{barnes-4-2}) because the series obtained from the calculation of the residues inside this closed contour is divergent for $y>1$ because the contour $C$
is closed to the left complex infinity for $y>1.$ Apparently, the only contribution from the contour $C$
corresponds to the left complex infinity which is equal to this divergent series according to the construction done
in Subsection \ref{MMG},
\begin{eqnarray*} 
 \frac{1}{2i\pi}  \oint\limits_{C} \frac{y^{-z}\le \psi(1) - \psi(1+z)  \ri}{z}dz  =   - \sum_{k=1}^{\infty}\frac{y^k}{k}  \\ 
=  \frac{1}{2i\pi}  \int\limits_{-\infty+i\infty}^{-\infty - i\infty}  \frac{y^{-z}\le \psi(1) - \psi(1+z)  \ri}{z}dz.
\end{eqnarray*}
By using the MB inverse transformation (\ref{moment-barnes-51-2}) for $y>1$ we may write for the same closed contour $C$ that
\begin{eqnarray*} 
F(y)  =  \ln(1-y) = \ln(-y) +\ln \le 1 - \frac{1}{y} \ri  \no \\
= \frac{1}{2i\pi}   \int\limits_{-1+\delta-i\infty}^{-1+\delta + i\infty}\frac{(-y)^{-z}\G(z)\G(1-z)}{z}  dz   \no\\
= \frac{1}{2i\pi}   \oint\limits_{C}\frac{(-y)^{-z}\G(z)\G(1-z)}{z}  dz \no\\
- \frac{1}{2i\pi} \int\limits_{-\infty+i\infty}^{-\infty - i\infty} \frac{(-y)^{-z}\G(z)\G(1-z)}{z}  dz   \no\\
  =  - \sum_{k=1}^{\infty}\frac{y^k}{k}  -   \frac{1}{2i\pi} \int\limits_{-\infty+i\infty}^{-\infty - i\infty} \frac{(-y)^{-z}\G(z)\G(1-z)}{z}  dz.
\end{eqnarray*}

Thus, we may write for the divergent series (when $y>1$) this chain of identities
\begin{eqnarray*} 
  - \sum_{k=1}^{\infty}\frac{y^k}{k}  = \frac{1}{2i\pi}  \int\limits_{-\infty+i\infty}^{-\infty - i\infty}  \frac{y^{-z}\le \psi(1) - \psi(1+z)  \ri}{z}dz \no\\
  =  \ln(1-y)  + \frac{1}{2i\pi} \int\limits_{-\infty+i\infty}^{-\infty - i\infty} \frac{(-y)^{-z}\G(z)\G(1-z)}{z}  dz.
  \end{eqnarray*}

Thus, even if we work with the divergent series and integrals  at the intermediate steps, we come to a correct relation
for the finite integrals
\begin{eqnarray} \label{y>12-2} 
\ln(1-y) = \frac{1}{2i\pi}  \int\limits_{-\infty+i\infty}^{-\infty - i\infty}  \frac{y^{-z}\le \psi(1) - \psi(1+z)  \ri}{z} \no\\
- \frac{1}{2i\pi} \int\limits_{-\infty+i\infty}^{-\infty - i\infty} \frac{(-y)^{-z}\G(z)\G(1-z)}{z}  dz \no\\
= \frac{1}{2i\pi}  \int\limits_{-\infty+i\infty}^{-\infty - i\infty} \frac{y^{-z}}{z}\le \psi(1)    - \psi(1+z) \right.   \no\\
 \left. -(-1)^{-z}\G(z)\G(1-z)\ri dz,
\end{eqnarray}
which  is valid for any $y \in [1,\infty[.$ At the same time, for $y \in [0,1]$ this integral (\ref{y>12-2})
is zero by construction. We may re-write the construction (\ref{y>12-2}) as a contour integral
\begin{eqnarray} \label{C4}
\ln(1-y) = \frac{1}{2i\pi}  \oint\limits_{C_4} \frac{y^{-z}}{z}\le \psi(1)    - \psi(1+z) \right.   \no\\
 \left. -(-1)^{-z}\G(z)\G(1-z)\ri dz \no\\
 = \frac{1}{2i\pi}  \int\limits_{-\infty+i\infty}^{-\infty - i\infty} \frac{y^{-z}}{z}\le \psi(1)    - \psi(1+z) \right.   \no\\
 \left. -(-1)^{-z}\G(z)\G(1-z)\ri dz,
\end{eqnarray}
such that all the poles of both Euler gamma functions $\G(z)$ and  $\G\le 1 -z\ri$ are inside
the contour $C_4.$ This is the only condition for the contour $C_4.$ However, the contour integral representation (\ref{C4}) is valid only for $ y > 1$. We cannot apply the contour representation   (\ref{C4})
for  $y \in [0,1]$ because a divergent series would appear on the r.h.s. but the vertical line in (\ref{y>12-2})
may be supplemented by the vertical line of (\ref{killing-22}) in order to  close a rectangular contour
and the result for the integral would remain zero for such a  closed contour for the standard domain $y \in [0,1].$

At the same time, it is clear that
in  (\ref{killing-22-2}) the contour $C$ may be taken as this contour $C_4,$ too, because the same residues remain inside it.  Thus, for any $y \in [0,1]$
we may write
\begin{eqnarray} \label{y<12-2}
\ln(1-y)  =  \frac{1}{2i\pi}  \oint\limits_{C_4}\frac{y^{-z}}{z}\le \psi(1)  - \psi(1+z)  \ri dz \no\\
= \frac{1}{2i\pi}  \int\limits_{-1+\delta-i\infty}^{-1+\delta + i\infty} \frac{y^{-z}\le \psi(1) - \psi(1+z)  \ri}{z}dz.
\end{eqnarray}
The advantage of the representation in terms of a contour integral is that  $C_4$ is a closed contour from the very beginning but not an open line
like in the MB transformation (\ref{moment-barnes-51-2}).

Another useful representation may be obtained by combining   (\ref{y<12-2}) and (\ref{C4}),
\begin{eqnarray} \label{any}
\ln(1-y) =  \frac{1}{2i\pi}  \int\limits_{-\infty+i\infty}^{-\infty - i\infty} \frac{y^{-z}}{z}\le \psi(1)    - \psi(1+z) \right.   \no\\
 \left. -(-1)^{-z}\G(z)\G(1-z)\ri dz \no\\
+ \frac{1}{2i\pi}  \int\limits_{-1+\delta-i\infty}^{-1+\delta + i\infty} \frac{y^{-z}\le \psi(1) - \psi(1+z)  \ri}{z}dz \no\\
= \frac{1}{2i\pi}  \oint\limits_{C_4}\frac{y^{-z}\le \psi(1)  - \psi(1+z)  \ri}{z} dz \no\\
- \frac{1}{2i\pi}  \int\limits_{-\infty+i\infty}^{-\infty - i\infty} \frac{(-y)^{-z} \G(z)\G(1-z)}{z} dz.
\end{eqnarray}
This representation  (\ref{any}) is valid for any $y \in [0,\infty[.$

\subsection{Example of Laplace transform with infinitely many poles} \label{Sub-Lap}

We wrote at the beginning of Section \ref{MMB} that the relation between the Laplace transform and the Mellin moment is straightforward. Their arguments may be related by a simple change of variables (\ref{MMT Cauchy}). One of these transformations is just a copy of another. In this Subsection \ref{Sub-Lap} we consider an example of the transforming function $f(x)$ for which the radius of convergence is limited, $f(x) = \ln(1-e^{-x}).$ This example is closely related to the moment of the function  $F(y) = \ln(1-y)$ considered in the previous Subsection \ref{Sub-psi}.
All the formulae of this Subsection would be copies of the corresponding formulae of the previous Subsection
\ref{Sub-psi}. We simply omit most of them remaining with the copies of the first formula (\ref{moment-barnes-50-2}) and of the last two formulas (\ref{C4}) and (\ref{y<12-2}).

We come to this expression for the Laplace transform of $f(x) = \ln(1-e^{-x})$ starting from
the Mellin moment of the function  $F(y) = \ln(1-y)$, which may be expressed via the Euler digamma function,
\begin{eqnarray} \label{change}
\int\limits_{0}^{1} y^{z-1}\ln{(1-y)}~ dy = \int\limits_{0}^{\infty} e^{-zx}\ln{(1-e^{-x})}~ dx\no\\
= \frac{\psi(1) - \psi(1+z) }{ z}.
\end{eqnarray}

The standard inverse integral transformation (\ref{MMT Cauchy}) for the Laplace transform recovers the original function
\begin{eqnarray}\label {moment-barnes-50-3}
 \frac{1}{2i\pi}  \int\limits_{\delta-i\infty}^{\delta + i\infty} \frac{e^{xz}\le \psi(1) - \psi(1+z)  \ri}{z}dz   =  \ln{(1-e^{-x})}.
\end{eqnarray}
This formula  (\ref{moment-barnes-50-3})  is just a copy of the formula  (\ref{moment-barnes-50-2}) of the previous Subsection \ref{Sub-psi}.

Repeating all the formula of Subsection \ref{Sub-psi} with a minimal modification
(\ref{change}) we come to the analog  of its  final result (\ref{C4}), that is

\begin{eqnarray}
\ln(1-e^{-x}) = \frac{1}{2i\pi}  \oint\limits_{C_4} \frac{e^{xz}}{z}\le \psi(1)    - \psi(1+z) \right.   \no\\
 \left. -(-1)^{z}\G(z)\G(1-z)\ri dz \label{C4-2}\\
 = \frac{1}{2i\pi}  \int\limits_{-\infty+i\infty}^{-\infty - i\infty} \frac{e^{xz}}{z}\le \psi(1)    - \psi(1+z) \right.   \no\\
 \left. -(-1)^{z}\G(z)\G(1-z)\ri dz. \label{line}
\end{eqnarray}
This contour integral representation (\ref{C4-2}) is valid only for $ x < 0.$ We cannot apply the contour representation   (\ref{C4-2})
for  $x \in [0,\infty[$ because a divergent series would appear on the r.h.s. of the vertical line in (\ref{line}).
However, the representation with the open straight vertical line (\ref{line}) at the left complex infinity may be applied for any $x \in ]-\infty,\infty[$ and recovers zero for  $x \in [0,\infty[.$

An analog of the formula (\ref{y<12-2}) for the Laplace transformation would be
\begin{eqnarray} \label{y-222}
\ln(1-e^{-x})  =  \frac{1}{2i\pi}  \oint\limits_{C_4}\frac{e^{xz}}{z}\le \psi(1)  - \psi(1+z)  \ri dz \no\\
= \frac{1}{2i\pi}  \int\limits_{-1+\delta-i\infty}^{-1+\delta + i\infty} \frac{e^{xz}\le \psi(1) - \psi(1+z)  \ri}{z}dz.
\end{eqnarray}
This formula is valid for any $x$ in the standard domain $x \in [0,\infty[.$

Finally, we write down a formula for the inverse Laplace transformation which is valid for any real
$x \in ]-\infty,\infty[,$
\begin{eqnarray} \label{any-2}
\ln(1-e^{-x}) =  \frac{1}{2i\pi}  \int\limits_{-\infty+i\infty}^{-\infty - i\infty} \frac{e^{xz}}{z}\le \psi(1)    - \psi(1+z) \right.   \no\\
 \left. -(-1)^{z}\G(z)\G(1-z)\ri dz \no\\
+ \frac{1}{2i\pi}  \int\limits_{-1+\delta-i\infty}^{-1+\delta + i\infty} \frac{e^{xz}\le \psi(1) - \psi(1+z)  \ri}{z}dz \no\\
= \frac{1}{2i\pi}  \oint\limits_{C_4}\frac{e^{xz}\le \psi(1)  - \psi(1+z)  \ri}{z} dz \no\\
- \frac{1}{2i\pi}  \int\limits_{-\infty+i\infty}^{-\infty - i\infty} \frac{e^{xz} (-1)^z   \G(z)\G(1-z)}{z} dz.
\end{eqnarray}

\section{Bethe-Salpeter Equations  for structure functions of proton} \label{Sec-pro}

In this Section we briefly review how the Bethe-Salpeter equations of quantum chromodynamics can be re-written as equations of quantum mechanics.  The Bethe-Salpeter equations were used  by Gribov and Lipatov in quantum electrodynamics to study the structure functions   of proton in the framework of a parton model \cite{Gribov:1972ri,Gribov:1972rt,Lipatov:1974qm}.  These IDEs  appeared at the  level of Feynman diagrams  to describe the processes of   deeply inelastic lepton scattering  in which proton structure functions are measured  \cite{Gribov:1972ri,Gribov:1972rt,Lipatov:1974qm}.  Then,  operator product expansion in combination with the  later discovered  asymptotic freedom for the running coupling in non-Abelian gauge theories were applied to construct and solve the evolution equation for the proton structure functions  \cite{Gross:1974cs}. There are many review papers dedicated to  these scattering processes, we mainly refer  to   \cite{Collins:1989gx}.

\subsection{DGLAP equation}

The proton structure functions are measured experimentally \cite{Collins:1989gx}. 
The standard hadronic tensor $W^{\mu\nu} (q^\mu, p^\mu)$,
\begin{eqnarray} \label{W}
  &&W^{\mu\nu}  = \frac{1 }{4\pi} \int d^4y e^{ i q \cdot y}
      \sum_X  \langle A|j ^\mu (y) | X \rangle \langle X|j^\nu(0) | A \rangle
\nonumber\\
  &=& F_1(x, Q^2)  \left( -g^{ \mu\nu} + \frac{q^\mu q^\nu }{q^2} \right)
\nonumber\\
  & +& F_2(x, Q^2) \frac{ \left( p^\mu - q^\mu p \cdot q / q^2 \right)
                         \left( p^\nu - q^\nu p \cdot q / q^2 \right)}{p \cdot q}
\end{eqnarray}
may be extracted from measurements in deeply inelastic lepton scattering process
$e + A \to e + X$, see \cite{Collins:1989gx}, in which  $q^\mu$ is the momentum of
a virtual photon of the exchange, $Q^2 = - q_\mu q^\mu$,
$x = Q^2 / 2 q \cdot p$, $p^\mu$ is the momentum of
the incoming hadron $A$, and $j^\mu(y)$ is the electromagnetic current.  The vectors $q^\mu, j^\mu(y), p^\mu$ are vectors in the four-dimensional Minkowski space.

According to the parton model, the pointlike partons were ``seen'' inside the hadron in the scattering experiments.  In the framework of this approach,  the following  integral  representation is used for the proton structure functions, 
\begin{eqnarray}  
 F_1\left(x, \frac{Q^2}{\Lambda^2}\right)   = \sum_{a}\int\limits_x^1 \frac{d \xi}{\xi}  \ f_{a/A}\left(\xi, \frac{\mu^2}{\Lambda^2} \right)   \nonumber \\
H_{1a} \left( \frac{x}{\xi}, \frac{Q^2}{\mu^2}, \alpha\left(\frac{Q^2}{\mu^2}\right)\right),  \label{F1}  \\ 
  \frac{1 }{x}F_2\left(x, \frac{Q^2}{\Lambda^2}\right)     = \sum_{a}
      \int\limits_x^1 \frac{d \xi}{\xi}  \ f_{a/A}\left(\xi, \frac{\mu^2}{\Lambda^2} \right)  \frac{\xi }{x} \nonumber \\
H_{2a} \left( \frac{x}{\xi}, \frac{Q^2}{\mu^2},  \alpha\left(\frac{Q^2}{\mu^2}\right)\right).  \label{F2}  
\end{eqnarray}
Here $ f_{a/A}\left(\xi, \frac{\mu^2}{\Lambda^2} \right)$ is a parton distribution function (PDF), $\mu^2$ is a renormalization scale, $\Lambda^2$ is 
so-called QCD scale. The PDF $ f_{a/A}\left(\xi, \frac{\mu^2}{\Lambda^2} \right)$
means the probability to find a parton of
type $a $ in a hadron of type $A$ carrying a fraction $\xi$ to $\xi + d \xi$
of the hadron's momentum.  In the formula, one sums over all the possible
types of parton  $a$  \cite{Collins:1989gx}.  The functions $H_{1a},  H_{2a}$  may be calculated 
perturbatively.

The Mellin convolution of  the integrands of Eqs. (\ref{F1}) and    (\ref{F2}) has its origin in the factorization of the Mellin moments 
of the proton structure functions, 
\begin{eqnarray}   \label{mome}
   \tilde F_1\left(N, \frac{Q^2}{\Lambda^2}\right)
   &=& \int\limits_0^1 \frac{d x}{x}\ x^N F_1\left(x,\frac{Q^2}{\Lambda^2}\right) , \nonumber\\
   \tilde F_2\left(N, \frac{Q^2}{\Lambda^2}\right)
   &=& \int\limits_0^1 \frac{d x}{x} \ x^{N-1} F_2 \left(x, \frac{Q^2}{\Lambda^2}\right),   \nonumber\\
\tilde f_{a/A}\left(N, \frac{\mu^2}{\Lambda^2} \right)
   &=& \int\limits_0^1 \frac{d x}{x} \ x^{N} f_{a/A}\left(x, \frac{\mu^2}{\Lambda^2} \right),
\end{eqnarray}
which become sums of    products of the Mellin moments,
\begin{eqnarray}
\tilde F_i \left(N, \frac{Q^2}{\Lambda^2}\right) \nonumber\\
= \sum_{a}
 \tilde f_{a/A}\left(N, \frac{\mu^2}{\Lambda^2} \right)
\tilde H_{i a} \!\!\left( N, \frac{Q^2}{\mu^2}, \alpha\left(\frac{Q^2}{\mu^2}\right)\right).   \label{start}
\end{eqnarray}

The structure of  (\ref{start}), factorized in the product of the perturbative part and the PDFs, comes from the solution to the renormalization group equations (RGEs)  for products of operators  \cite{Gross:1974cs}.      Initially, at some scale,  the Mellin moment  of the proton structure function  is taken as  a product of a factor depending on the QCD scale  $\Lambda$  and on the  Mellin variable $N,$  and  the Mellin moment of a coefficient  function \cite{Gross:1974cs}.  The latter may be calculated in a perturbative way \cite{Collins:1989gx}.  The moments of PDFs are parametrized at certain scale in certain way, for example,  by using the Euler beta function \cite{Alekhin:2003ev}.   At another scale they are different and evolve according  to  the RGE.

Thus,  the product of the perturbative part and the PDF should satisfy the evolution equations  with respect to momentum transfer  $Q^2,$  which may be transformed to the RGE with respect to $Q^2$   for the  perturbative part $\tilde H_{i a} \!\!\left( N, \frac{Q^2}{\mu^2}, \alpha\left(\frac{Q^2}{\mu^2}\right)\right).$   The l.h.s. of 
(\ref{start})  does not depend on the renormalization scale $\mu.$ This means that the RGE for the perturbative part 
$\tilde H_{i a} \!\!\left( N, \frac{Q^2}{\mu^2}, \alpha\left(\frac{Q^2}{\mu^2}\right)\right)$  of (\ref{start}) may be transformed to the RGE for the Mellin moments $\tilde f_{a/A}\left(N, \frac{\mu^2}{\Lambda^2} \right)$   of the  PDFs.   Each factor  in (\ref{start}) satisfies  its own evolution  equation with respect to  the renormalization group. However, these evolution equations are related because the l.h.s. does not depend on $\mu.$ In particular, the anomalous dimensions of the Mellin moments of the PDFs coincide with anomalous dimensions of the operators of highest twist from the OPE  \cite{Altarelli:1977zs}.

In the previous paragraph we wrote that the RGEs for the nonperturbative part and the perturbative parts of the proton structure functions are related. The PDFs are nonperturbative part of this decomposition.  The DGLAP IDE comes from the evolution equation for the structure functions,   it is the RGE for the PDFs with respect the  renormalization scale $\mu$.  
Let us recall a simple model considered in  \cite{Alvarez:2016juq}. It contains the main features of the  RGEs for the PDFs.  Because of  the reasons explained in \cite{Alvarez:2016juq} the dominant  partons  in the region of small $x$ are gluons (gauge bosons of QCD).  The sum over partons  in (\ref{start}) may be omitted in this approximation  and we leave with a gluon distribution function only, 
 \begin{equation}
\tilde F_i \left(N, \frac{Q^2}{\Lambda^2}\right) \sim   \tilde f \left(N, \frac{\mu^2}{\Lambda^2} \right)
\tilde H_{i} \left( N, \frac{Q^2}{\mu^2}, \alpha\left(\frac{Q^2}{\mu^2}\right)\right).    \label{domi}
\end{equation}
It was shown by  Altarelli and Parisi in \cite{Altarelli:1977zs}  that the evolution equations for the PDF is 
 \begin{eqnarray}  \label{rge}
u \frac{d}{d u} \tilde f(N,u)  = \frac{\alpha(u)}{2\pi}~\gamma(N,\alpha(u))~\tilde f(N,u),
\end{eqnarray}
where we have introduced the notation $u \equiv \mu^2/\Lambda^2,$ and $\gamma(N,\alpha(u))$ is the anomalous dimension of the gluon  distribution function.  The RGEs for the perturbative parts 
$\tilde H_{i} \left( N, \frac{Q^2}{\mu^2}, \alpha\left(\frac{Q^2}{\mu^2}\right)\right)$ 
are more complicate \cite{Gross:1974cs}.  The running coupling may be solved in a perturbative way for large momentum transfers, too \cite{Cvetic:2011vy}.  

The RGE for the  Mellin moments of the PDFs can be represented as an IDE for the same PDFs \cite{Altarelli:1977zs}.  After transforming  Mellin moments of a PDF back to the PDF we obtain  the DGLAP IDE  for the PDF
\begin{eqnarray}  
u\frac{d}{du} f(x,u) =  \frac{\alpha(u)}{2\pi}\int\limits_x^1\frac{dy}{y}~f(y,u)P\left(\frac{x}{y}, \alpha(u) \right),  \label{DGLAP-1} \\
\int\limits_0^1 dx ~x^{N-1} P\left(x, \alpha(u)\right) = \gamma(N,\alpha(u)), \label{split}
\end{eqnarray}
These equations for the PDF are similar to  Gribov and Lipatov IDEs for the PDFs  found in quantum electrodynamics (QED) based on Bethe-Salpeter  equation with similar  splitting functions   $P\left(\frac{x}{y}, \alpha(u)\right) $ in the integral kernels of these equations  
\cite{Gribov:1972ri,Gribov:1972rt,Lipatov:1974qm}.   The anomalous dimension  $\gamma(N,\alpha(u))$  is the Mellin moment  of the splitting function    $P\left(\frac{x}{y}, \alpha(u)\right). $ The results of Gribov and Lipatov \cite{Gribov:1972ri,Gribov:1972rt,Lipatov:1974qm} were generalised  from QED to QCD by Dokshitzer \cite{Dokshitzer:1977sg}.

The RGE for the running coupling is solved only  up to some order of the perturbation theory \cite{Cvetic:2011vy}.  The solution of \cite{Cvetic:2011vy} for the running coupling  shows the asymptotic freedom at large momentum transfers $Q^2.$  As to the region of low momentum transfers $Q^2,$   the running coupling  is not necessary  large or infinite \cite{Ayala:2017tco}.  However, there is one model  \cite{Novikov:1983uc,Jones} of ${\cal N}=1$ supersymmetric pure Yang-Mills  theory with  $SU(N)$ as a gauge group,   in which the running  coupling  may be solved analytically in all the orders of the perturbation theory    \cite{Cvetic:2011vy} . The solution given in   \cite{Cvetic:2011vy} is
\begin{eqnarray*}
\alpha\left(  \frac{Q^2}{\mu^2} \right)   = -\frac{2\pi}{N} \frac{1}{W_{\mp 1}\left(-(\mu^2/Q^2)^{3/2}\right)}
\end{eqnarray*}
in which $W_{\mp 1}$ are different branches of Lambert function on the complex plane.   In this formula $N$ is from the the gauge group $SU(N)$ of the Yang-Mills theory. In all the other formulas of this paper $N$ is the variable of the Mellin moment.  This solution shows asymptotic freedom at large momentum transfers   $Q^2$ \cite{Cvetic:2011vy}.

\subsection{From DGLAP equation to optic theorem and vice verse }

The RGEs  for the perturbative parts of the proton structure functions from Ref. \cite{Gross:1974cs}  correspond to the RGEs for the Green functions of the Bethe-Salpeter equations written in  \cite{Dokshitzer:1977sg}.   Solutions to the RGEs for the Green functions can be written in terms of the contour integral in the complex plane of the Mellin moments as well as   the solutions to the RGEs for the PDFs, which are  nonperturbative parts of the structure functions, were written as contour integrals.  On the other hand, such Green functions renormalized by the propagators   become invariant with respect to  the renormalization group \cite{Kondrashuk:2004pu}.  Such   RG  invariant constructions from the four-point Green functions and the propagators  can be represented in terms of contour integrals in the complex plane of the Mellin moments, too.   After complex mapping of such a  contour integral to a dual contour integral, the dual contour integral, being implicitly RG invariant by  construction, may solve explicitly another IDE.  For the on-shell values of the external momenta involved in the scattering process,  this second IDE generates two different IDEs.  One of them is the  optic theorem written for the amplitude of the scattering process while the other is  an IDE for the PDFs  such that in the Regge limit it becomes the Balitsky-Fadin-Kuraev-Lipatov   (BFKL) equation for the PDFs.  In the Regge limit only one pole of the BFKL eigenvalue in the corresponding complex plane is taken into account \cite{Kondrashuk:2019cwi}.

\subsection{Complex map to a dual contour integral}

The gluon distribution $f(x,u)$ in (\ref{DGLAP-1})  is called "integrated" gluon distribution. It is related to the "unintegrated" gluon distribution $\varphi(x,k_\perp^2)$ by the relation 
\begin{eqnarray*}
f\le x,\frac{Q^2}{\mu^2}\ri = \int_0^{Q^2}dk_\perp^2 \varphi\le x,\frac{k_\perp^2}{\mu^2}\ri. 
\end{eqnarray*}
It appears, it is always possible to construct from $\varphi(x,k_\perp^2)$ a function which satisfies the same DGLAP equations as well as the integrated gluon distributions do. 
For example, in case when the coupling does not run, we have the same DGLAP equation for  $ \phi\le N,\frac{Q^2}{\mu^2}\ri \equiv Q^2\varphi\le N,\frac{Q^2}{\mu^2}\ri,$
that is \cite{Alvarez:2016juq}
\begin{eqnarray*}
Q^2\frac{d}{dQ^2} Q^2\varphi\le N,\frac{Q^2}{\mu^2}\ri = \frac{\alpha}{2\pi}\gamma(N,\alpha)   Q^2  \varphi\le N,\frac{Q^2}{\mu^2}\ri.
\end{eqnarray*}
In the case when the coupling depends on the scale, a construction similar to $ \phi\le x,\frac{Q^2}{\mu^2}\ri$ 
exists (see  \cite{Alvarez:2016juq})  but it is not related to the unintegrated PDF $\varphi\le x,\frac{k_\perp^2}{\mu^2}\ri$  in such a straightforward  way like it is in this case when the coupling does not run.  The IDE dual to DGLAP IDE is written for the unintegrated PDF by the reason explained in Subsection \ref{BFKL-sec}.

Going back to the simple model  (\ref{rge}) in which the only  partons are gluons, we find that it can be  a good example to show how a solution to DGLAP IDE in terms of contour integrals appears.   In case when the coupling does not depend on the scale $u$ in Eq. (\ref{rge}), the solution of such an equation is just a power-like dependence on the variable $u.$   To find the solution to the DGLAP equation we need to solve the differential equation  (\ref{rge})   for the Mellin moments and then to take the inverse Mellin transformation by calculating the residues according to Cauchy integral formula (\cite{Alvarez:2016juq})  or to take it  via complex mapping   \cite{Alvarez:2019eaa}. We take a  model with an anomalous dimension $\gamma(N) = 1/(N+1)$ and make  integral transformations.

The solution to the DGLAP equation (\ref{DGLAP-1}) may be represented   in the form of the contour integral 
\begin{eqnarray} \label{orig}
 \phi(x,u) =  \frac{1}{2\pi i}\int\limits_{-1+\delta-i\infty}^{-1+\delta+i\infty}~dN \frac{x^{-N}}{N+1} u^{ \dis{ \frac{1}{N+1}}} \no \\
= xI_0\le 2\sqrt{\ln{u}\ln{\frac{1}{x}}}\ri.
\end{eqnarray} 
This result is obtained by calculating the residues. 
 Since $x \in [0,1]$ the inverse transformation from the Mellin moment to the Bessel function in (\ref{orig}) is standard and does not require the modification  (\ref{MCRG})  constructed in Subsection \ref{MMG}.

In this paragraph we construct the dual contour integral by complex mapping   and show that it brings the same  Bessel function  $xI_0\le 2\sqrt{\ln{u}\ln{\frac{1}{x}}}\ri$ as it is in (\ref{orig}).  The new complex  variable $M$  is related to the original complex variable $N$ by the duality relation $\gamma(N) = M. $  If we define the inverse function $\chi(M) \equiv N,$ then we model the duality condition  $\chi(\gamma(N)) =  N$ and $\gamma(\chi(M)) =  M.$   The complex map based on this condition was performed in \cite{Kondrashuk:2019cwi}.  We may write explicitly 
\begin{eqnarray*} 
\label{complex diffeo} \gamma(N) =  \frac{1}{N+1} = M \Rightarrow  N = \frac{1}{M} -1 \equiv \chi(M), \\  
\label{inverse} \chi(\gamma(N)) = \frac{1}{\gamma(N)} -1 = (N+1) -1 = N, \\
 \gamma(\chi(M)) = \frac{1}{\chi(M) +1} =  \frac{1}{\left(\displaystyle{\frac{1}{M}-1}\right)  +1} = M, 
\end{eqnarray*}
and make the corresponding diffeomorphism in the complex plane of the variable $N$
\begin{eqnarray*} 
\phi(x,u) =  \frac{1}{2\pi i}\int\limits_{-1+\delta-i\infty}^{-1+\delta+i\infty}~dN \frac{x^{-N}}{N+1} u^{ \displaystyle{ \frac{1}{N+1} }} \\
=  \frac{1}{2\pi i}\oint\limits_{C}~dM N'(M) \frac{x^{-\chi(M)}} {N+1} u^M = - \frac{1}{2\pi i}\oint\limits_{C}~dM \frac{x^{-\chi(M)} }{M} u^M \\
= -\frac{1}{2\pi i}\oint\limits_{C}~dM \frac{x^{{1-1/M}}}{M} u^M =  - \frac{x}{2\pi i}\oint\limits_{C}~dM \frac{x^{-1/M}}{M} u^M \\
 = - x\sum_{k=0}^{\infty}\frac{1}{k!}\le -\ln{x}\ri^k \oint\limits_C~dM \frac{u^M}{M^{k+1}}  \\
= x\sum_{k=0}^{\infty}\frac{1}{k!}\le-\ln{x}\ri^k \frac{(\ln{u})^{k}}{k!} \\
= x\sum_{k=0}^{\infty}\frac{(-1)^{k}}{(k!)^2} \le\ln{u}\ln{x}\ri^k = xI_0\le 2\sqrt{\ln{u}\ln{\frac{1}{x}}} \ri.
\end{eqnarray*}
We have reproduced the same Bessel function, as it should be because any integral does not depend on any change of the variable. We may write the gluon unintegrated distribution function in a dual form 
\begin{eqnarray*} 
\phi(x,u) =  \frac{1}{2\pi i}\oint\limits_{C_1}~dN \phi(N,u) x^{-N} =  \frac{1}{2\pi i}  \oint\limits_{C_2}~dM \phi(x,M) u^M \\
= xI_0\le 2\sqrt{\ln{u}\ln{\frac{1}{x}}} \ri.
\end{eqnarray*}
The contour $C_2$ is an image of the contour $C_1$ with respect to the mapping in the complex plane of the Mellin moment.  The functions $\phi(N,u)$ and $\phi(x,M)$ satisfy to the different differential equations, 
\begin{eqnarray} 
u \frac{d}{d u} \phi \le N, u\ri   = \gamma(N,\alpha) \phi\le N, u\ri,  \label{eq1} \\
 x \frac{d}{d x} \phi \le x, M\ri   =  -\chi(M,\alpha) \phi\le x, M\ri.   \label{eq2}
\end{eqnarray}
The solutions to these equations are power-like functions, $\phi \le N, u \ri = \phi_1(N) u^{\gamma(N)}, ~~ \phi \le x, M \ri = \phi_2(M) x^{-\chi(M)},$ in which the functions  $\phi_1(N)$ and   $\phi_2(M)$ 
are related by the complex diffeomorphism (\ref{complex diffeo}) which is governed by the duality condition $\chi(\gamma(N)) =  N.$

The  DGLAP IDE (\ref{DGLAP-1})  is valid for   $ \phi\le x,u\ri$,  too  \cite{Alvarez:2016juq}.  It may be solved by the first contour integral
\begin{eqnarray}  \label{DGLAP-explicit}
\phi(x,u) = \frac{1}{2\pi i} \oint\limits_{C_1}~dN \phi(N,u) x^{-N}. 
\end{eqnarray}
 When we write this contour integral instead of $ \phi\le x,u\ri$  in  Eq. (\ref{DGLAP-1}), we obtain then the fist differential equation 
  (\ref{eq1}).  The dual contour integral 
\begin{eqnarray} \label{explicit} 
 \oint\limits_{C_2}~dM \varphi(x,M) u^M   \sim  \oint\limits_{C_2}~dM \phi(x,M) u^{M-1}  
\end{eqnarray}
solves another IDE explicitly. This is why we call this second IDE as a "dual DGLAP" IDE.  When we write this dual contour integral  (\ref{explicit}) in the dual DGLAP IDE we come to the second differential equation (\ref{eq2}).  That second IDE comes from the unitarity of the scattering matrix and corresponds to the optic theorem. The second IDE is constructed in Subsection  \ref{BFKL-sec}.

\subsection{BFKL equation} \label{BFKL-sec}

The  BFKL IDE was constructed in Refs.  \cite{Lipatov:1976zz,Fadin:1975cb,Kuraev:1976ge,Kuraev:1977fs,Balitsky:1978ic}. Let us start from the IDE for the Green function for the BFKL kernel.  It appears as a  four-point Green function of the reggeized gluons  in the Regge limit $s/{Q^2} \rightarrow \infty$  and   may be written in the form given in Eq. (9.113) of Ref. \cite{Ioffe:2010zz},
\begin{eqnarray} \label{BFKL-2}
\frac{d}{d Y}G(Y;q_{1\perp},q'_{1\perp};q_\perp) \no  \\ 
= \int\frac{d^{D-2}r_\perp}{r_\perp^2(q_\perp-r_\perp)^2}\text{\emph{K}}(q_{1\perp},r_\perp;q_\perp)G(Y;r_\perp,q'_{1\perp};q_\perp),
\end{eqnarray}
in which $Y$ is rapidity, $G(Y;q_{1\perp},q'_{1\perp};q_\perp)$ is a Green function of two reggeized gluons, the theory is dimensionally regularized, $D$ is a dimension of spacetime, 
$K(q_{1\perp}, r_\perp ; q_\perp)$ is a kernel of the BFKL equation, $q_{1\perp},q_{2\perp}$ are momenta of two incoming reggeized gluons, 
$q'_{1\perp},q'_{2\perp}$ are momenta of two outgoing reggeized gluons, and 
$ q_{\perp} = q_{1\perp} + q_{2\perp} = q'_{1\perp} + q'_{2\perp}$ is a transversal two-dimensional momentum in the $t$-channel of the ladder 
of reggeized gluons. The explicit form of the kernel $K(q_{1\perp}, r_\perp ; q_\perp)$ is given in Ref. \cite{Kotikov:2000pm}.

The rapidity is $Y=\ln(s/s_0)$ in which $s_0 = q_{1\perp}q'_{1\perp}$ of the Green function $G(Y; q_{1\perp}, q'_{1\perp}; q_\perp)$ is a scalar  product of two-dimensional momenta  $q_{1\perp}$ and $q'_{1\perp}$ 
On the other side, the Bjorken variable $x = Q^2/2pq$, where $p$ is a proton momentum and $q$ is a virtual photon momentum, $Q^2 = -q^2$  is the momentum transfer between electron and proton. 
According to the kinematics of the physical picture $s = (p+q)^2,$   
and $2pq = s - m^2 + Q^2,$ where $m$ is the proton mass. We consider the Regge kinematics in which $s$ is large, $s \gg Q^2,$ 
the only variable of the function $G(Y;q_{1\perp},q'_{1\perp};q_\perp)$ which depends on $s$ is $Y,$ and we may write
\begin{eqnarray*}
\frac{d}{dY}G(Y;q_{1\perp},q'_{1\perp};q_\perp) \backsimeq -x\frac{d}{dx}G(Y(x);q_{1\perp},q'_{1\perp};q_\perp). 
\end{eqnarray*}

It is supposed that all the dependence on the Mandelstam variable $s$  in the notation of the Green function $G(Y;q_{1\perp},q'_{1\perp};q_\perp)$ is
contained in the rapidity $Y$ while the dependence on $q_{1\perp},q'_{1\perp};q_\perp$ enters in this Green function via possible Lorentz scalar combinations which do not depend on $s.$ 
Under the condition that $q_{\perp} = 0$ for the forward scattering of the 
reggeized gluons we consider  $q_{1\perp}$ and $q'_{1\perp}$ as the momenta in the impact plane on which the impact factors of the initial 
and of the final states depend.   To evaluate the cross section  of the 
forward scattering, these impact factors should be convoluted via these momenta with $G(Y; q_{1\perp}, q'_{1\perp}; 0)$ \cite{Ioffe:2010zz},

Thus, we may rewrite IDE for the Green function (\ref{BFKL-2}) in the Regge limit for the Green function $G_R(x;q_{1\perp},q'_{1\perp};q_\perp)$ defined as  
$G_R(x;q_{1\perp},q'_{1\perp};q_\perp) \equiv  G(Y(x);q_{1\perp},q'_{1\perp};q_\perp),$
\begin{eqnarray} \label{BFKL-x-1}
-x\frac{d}{d x}G_R(x;q_{1\perp},q'_{1\perp};q_\perp) \no\\
=\int\frac{d^{D-2}r_\perp}{r_\perp^2(q_\perp-r_\perp)^2}\text{\emph{K}}(q_{1\perp},r_\perp;q_\perp)G_R(x;r_\perp,q'_{1\perp};q_\perp).
\end{eqnarray}
In order to use the Bethe-Salpeter equation for the Green function in the form of the optic theorem from Ref. \cite{Kuraev:1976ge}, we need to consider the forward scattering 
of two reggeized gluons, that is,
the  transversal two-dimensional momentum  $q_\perp$  in the $t$ channel of the ladder of reggeized gluons is zero, $q_\perp = 0,$
\begin{eqnarray} \label{BFKL-x-2}
-x\frac{d}{d x}G_R(x;q_{1\perp},q'_{1\perp};0) \no\\
=\int\frac{d^{D-2}r_\perp}{\le r_\perp^2\ri^2}\text{\emph{K}}(q_{1\perp},r_\perp;0)G_R(x;r_\perp,q'_{1\perp};0).
\end{eqnarray}

We may redefine the Green functions and the integral kernel  as 
\begin{eqnarray*}
G_R(x;q_{1\perp},q'_{1\perp};q_{\perp} ) =  q_{1\perp}^2q_{2\perp}^2\tilde{G}_R(x;q_{1\perp},q'_{1\perp};q_{\perp} ),  \\
K(q_{1\perp},q'_{1\perp};q_{\perp} ) =  q_{1\perp}^2q_{2\perp}^2\tilde{K}(q_{1\perp},q'_{1\perp};q_{\perp} ), \\
G_R(x;q_{1\perp},q'_{1\perp};0) =  \le q_{1\perp}^2\ri^2\tilde{G}_R(x;q_{1\perp},q'_{1\perp};0 ), \\
K(q_{1\perp},q'_{1\perp};0 ) =   \le q_{1\perp}^2\ri^2 \tilde{K}(q_{1\perp},q'_{1\perp};0  ). 
\end{eqnarray*}
These redefined Green function $\tilde{G}_R$  and the redefined kernel $\tilde{K}$ have the dimension of Dirac $\delta$-function in the $(D-2)$-dimensional 
space \cite{Ioffe:2010zz}.  In terms of them Eq. (\ref{BFKL-x-2}) takes a form 
\begin{eqnarray} \label{BFKL-x-3}
-x\frac{d}{d x}\tilde{G}_R(x;q_{1\perp},q'_{1\perp};0) \no\\
=\int d^{D-2}r_\perp \tilde{K}(q_{1\perp},r_\perp;0)\tilde{G}_R(x;r_\perp,q'_{1\perp};0).
\end{eqnarray}
We may replace in the  IDE   (\ref{BFKL-x-3})   the Green function $\tilde{G}_R(x;q_{1\perp},q'_{1\perp};0)$  with 
the inverse integral transformation of its Mellin moment   $\tilde{G}_R(N;q_{1\perp},q'_{1\perp};0)$ with respect to the Bjorken variable $x,$ constructed in  Subsection     \ref{MMG}, 
\begin{eqnarray} \label{IIT}
\tilde{G}_R(x; q_{1\perp},q'_{1\perp}; 0) \no\\
 =  \frac{1}{2\pi i}\oint\limits_{C_3}{d N}~x^{-N}~\tilde{G}_R(N; q_{1\perp},q'_{1\perp};0), 
\end{eqnarray}
where  
\begin{eqnarray*}
\tilde{G}_R(N;q_{1\perp},q'_{1\perp};0)  
= \int\limits_0^{1}~dx~x^{N-1}~\tilde{G}_R(x;q_{1\perp},q'_{1\perp};0).
\end{eqnarray*}
All the poles of the function $\tilde{G}_R(N;q_{1\perp},q'_{1\perp};0)$ in the complex plane of the variable $N$ 
are inside  the contour $C_3$  (see Subsection     \ref{MMG}).

After replacing the Green function in Eq. (\ref{BFKL-x-3}) with the inverse integral transformation (\ref{IIT}) we come to 
\begin{eqnarray} \label{BFKL-x-4}
N\tilde{G}_R(N;q_{1\perp},q'_{1\perp};0) \no \\
=\int d^{D-2}r_\perp \tilde{K}(q_{1\perp},r_\perp;0)\tilde{G}_R(N;r_\perp,q'_{1\perp};0).
\end{eqnarray}
The Green equation written in such a form means that the Green function can be expressed in terms of eigenfunction of the integral operator on the right hand side. The eigenvalue is $N.$  However, the eigenfunction and the eigenvalue of this integral operator are  well-studied. They may be found in    \cite{Ioffe:2010zz} and  \cite{Kotikov:2000pm}, 
\begin{eqnarray} \label{EF} 
\int d^{D-2}r_\perp \tilde{K}(q_{1\perp},r_\perp;0)\le\frac{r_{\perp}^2}{\mu^2}\ri^{\gamma - 1} = \chi(\gamma)  \le\frac{q_{1\perp}^2}{\mu^2}\ri^{\gamma - 1},
\end{eqnarray}
the function $\chi(\gamma)$ may be found in   Refs. \cite{Ioffe:2010zz} and  \cite{Kotikov:2000pm}, too.  Here $\gamma$ is arbitrary complex number.

Now we search for the solution to the forward Green function $\tilde{G}_R(N;q_{1\perp},q'_{1\perp};0)$ of two reggeized gluons. Such a Green function should satisfy Eq.  (\ref{BFKL-x-4}). 
This Green function may be written in terms of unintegrated gluon distribution function. There is a simple argument in favor of this statement. Indeed, according to  Eq. (\ref{EF}) we may try this pattern    
\begin{eqnarray*} 
\tilde{G}_R(N;q_{1\perp},q'_{1\perp};0) \sim  \le\frac{q_{1\perp}^2}{{q'}_{1\perp}^2}\ri^{g(N)-1},
\end{eqnarray*}
where $g(N)$ is a function of the complex variable $N.$ We did not fix this function yet. By putting it in Eq. (\ref{BFKL-x-4}) we observe that this function should be inverse to the eigenvalue $\chi(\gamma)$ of the integral operator (\ref{EF}),  
\begin{eqnarray*} 
N \le\frac{q_{1\perp}^2}{{q'}_{1\perp}^2}\ri^{g(N) - 1}   = \int d^{D-2}r_\perp \tilde{K}(q_{1\perp},r_\perp;0) \le\frac{r_{\perp}^2}{{q'}_{1\perp}^2}\ri^{g(N) - 1} \\
= {\chi}(g(N))\le\frac{q_{1\perp}^2}{{q'}_{1\perp}^2}\ri^{g(N) - 1}
\end{eqnarray*}
By the duality condition  $\chi({\gamma}(N)) = N$    this means $g(N) = {\gamma}(N).$  Vice verse, the duality condition comes from  structure of the integral kernel $\tilde{K}(q_{1\perp},r_\perp;0) $ and  its eigenfunction. 
 Then, taking into account the dimension of the Green function 
$\tilde{G}_R(N;q_{1\perp},q'_{1\perp};0)$  we may write as a solution 
\begin{eqnarray*} 
\tilde{G}_R(N;q_{1\perp},q'_{1\perp};0) =   \varphi\le N, \frac{q_{1\perp}^2}{{q'}_{1\perp}^2}\ri \sim 
\frac{1}{{q'}_{1\perp}^2}  \le\frac{q_{1\perp}^2}{{q'}_{1\perp}^2}\ri^{{\gamma}(N)-1}. 
\end{eqnarray*}

By construction,   this expression 
\begin{eqnarray*} 
\phi\le N, \frac{q_{1\perp}^2}{{q'}_{1\perp}^2}\ri  =   q_{1\perp}^2\varphi\le N, \frac{q_{1\perp}^2}{{q'}_{1\perp}^2}\ri
\sim \le\frac{q_{1\perp}^2}{{q'}_{1\perp}^2}\ri^{{\gamma}(N)} 
\end{eqnarray*} 
solves the DGLAP  IDE  (\ref{DGLAP-1}) in the form of the first contour integral (\ref{DGLAP-explicit}).  The dual contour integral 
(\ref{explicit})  solves explicitly another IDE (\ref{BFKL-x-3}) for the unintegrated distribution function 
\begin{eqnarray}  \label{BFKL-Lipatov} 
-x\frac{d}{d x}  \varphi\le x, \frac{q_{1\perp}^2}{{q'}_{1\perp}^2}\ri \no\\
=\int d^{D-2}r_\perp \tilde{K}(q_{1\perp},r_\perp;0) \varphi\le x, \frac{r_{\perp}^2}{{q'}_{1\perp}^2}\ri.  
\end{eqnarray}
Indeed, replacing  $\varphi\le x, \frac{q_{1\perp}^2}{{q'}_{1\perp}^2}\ri$ with the dual integral (\ref{explicit})
\begin{eqnarray}\label{DIR} 
 \oint\limits_{C_2}~dM \varphi(x,M) u^M   \sim  \oint\limits_{C_2}~dM \phi(x,M) u^{M-1} \no\\
=    \oint\limits_{C_2}~dM \phi(M)x^{-\chi(M)} u^{M-1},
\end{eqnarray}
in which $\phi(M)$ is a function of the complex variable $M,$ we observe  from  (\ref{BFKL-Lipatov})  that  
\begin{eqnarray*}  
-x\frac{d}{d x}  \oint\limits_{C_2}~dM \phi(M)x^{-\chi(M)}   \le\frac{q_{1\perp}^2}{{q'}_{1\perp}^2}\ri^{M-1}   \no\\
=\int d^{D-2}r_\perp \tilde{K}(q_{1\perp},r_\perp;0) \oint\limits_{C_2}dM \phi(M)x^{-\chi(M)}   \le\frac{r_{\perp}^2}{{q'}_{1\perp}^2}\ri  ^{M-1} \\
=  \oint\limits_{C_2}dM \phi(M)x^{-\chi(M)} \chi(M)   \le\frac{q_{1\perp}^2}{{q'}_{1\perp}^2}\ri^{M-1}.
\end{eqnarray*}
 Thus, the dual contour integral (\ref{DIR}) solves explicitly the dual DGLAP IDE (\ref{BFKL-Lipatov}). The dual contour $C_2$ is constructed in Subsection    \ref{MMG} exactly for this case when the variable $u \in [0,\infty[.$   This IDE  (\ref{BFKL-Lipatov})  is a consequence of the optic theorem considered in the Regge limit and it is called Balitsky-Fadin-Kuraev-Lipatov (BFKL) equation.

\section{Conclusions}

To analyze quantum communications between quantum computers, we need to solve the Schrodinger equation for the corresponding quantum systems \cite{Kondrashuk:2025ajb}.  The optic theorem may be written as a Schrodinger equation because its Regge limit, which is called the BFKL equation \cite{Fadin:1975cb,Kuraev:1976ge,Kuraev:1977fs,Balitsky:1978ic},  has already been written as a 
Schrodinger equation in  \cite{Lipatov:1993yb}.   

The proton structure functions in QCD may be studied by means of operator product expansion  \cite{Gross:1974cs}.  
There are quantum field theories  in which due to different reasons    the gauge coupling does not depend on the scale of the scattering process  \cite{Avdeev:1992jt,Kazakov:1995cy,Kondrashuk:1997uf,Alvarez:2016juq,Kondrashuk:2004pu}.    In such theories the operator product expansion  may be applied for any distances  \cite{Kondrashuk:2025ajb}. 
    The DGLAP  equation which is the RGE for the Mellin moments of the proton structure functions
\cite{Gribov:1972ri,Gribov:1972rt,Lipatov:1974qm,Lipatov:1976zz,Dokshitzer:1977sg,Altarelli:1977zs}  may be solved in terms of a  contour integral in the complex plane of the Mellin moments. Such a  contour integral may be transformed to a dual contour integral via complex mapping    \cite{Kondrashuk:2019cwi,Alvarez:2019eaa}.
That dual integral in turn solves the optic theorem in such scale-independent theories  \cite{Kondrashuk:2019cwi,Alvarez:2019eaa,Kondrashuk:2025ajb}.  The optic theorem is a consequence of  unitarity of the scattering matrix in  quantum field theory.  Vice verse, by solving the optic theorem in terms of a contour integral and then by transforming it to a dual contour integral we may construct the corresponding   RGE such that  the dual contour integral solves this constructed equation. 

 Starting with the optic theorem for the theories with the running coupling, we may solve it in terms of the contour integral, make a complex mapping to a dual contour integral and find a normalization group  equation whose solution is the dual integral.  By construction this RGE in the perturbative region 
should coincide with the DGLAP  equation \cite{Cvetic:2011vy, Ayala:2017tco}.  The original DGLAP equation in the realistic theories with the running coupling like quantum 
chromodynamics  is valid only for large momentum transfer \cite{Gribov:1972ri,Gribov:1972rt,Lipatov:1974qm,Lipatov:1976zz,Dokshitzer:1977sg,Altarelli:1977zs}. 
The RGE dual to the optic theorem may be written as a Schrodinger equation even in the theories with the running gauge coupling as well as  the original DGLAP equation has been written as a Schrodinger equation by Lipatov in \cite{Lipatov:1974qm}. 

 Finally, the solution to the Schrodinger equation which may be obtained from the optic theorem is written  in terms  of a contour integral  \cite{Alvarez:2019eaa}
 because the solution to the dual RGE may be found in terms of the contour integral  \cite{Alvarez:2016juq,Kondrashuk:2019cwi,Alvarez:2019eaa,Kondrashuk:2025ajb}.  This observation  opens doors to efficient construction of the protocols for quantum communications in future quantum computers.   The contour integral may be solved via complex mapping in the plane of the complex moments by using Jacobians of the complex maps 
\cite{Alvarez:2019eaa,Kondrashuk:2025ajb}.

\section*{Acknowledgment}
All the contents of this paper  is based on the lectures on Differential Equations which I.K. gave at the Campus  Fernando May, UBB, Chillán for informatics students in the years 2017-2025.  He is grateful to the authorities of the School of Informatics at the UBB, Chill\'an,  for steady support.

%\bibliographystyle{IEEEtran}
%\bibliography{bibtexReferencesFileName}

\end{document}